\documentstyle[12pt]{article}
\newcommand{\nc}{\newcommand}
\nc{\rnc}{\renewcommand}
\rnc{\title}[1]{\Large\bf\mbox{}\\ \mbox{}\\#1\bigskip\medskip\\}
\rnc{\author}[1]{\large #1\\ \smallskip}
\nc{\address}[1]{{\narrower\normalsize\it #1\\}\medskip}
\nc{\e}[1]{{\em #1\/}}
\nc{\comment}[1]{}
\nc{\ADE}{$A$--$D$--$E$\ }
\nc{\prop}[1]{\noindent$\bullet$\ \ #1}
\rnc{\thesection}{\arabic{section}\,.}
\rnc{\thesubsection}{\arabic{section}.\arabic{subsection}}
\rnc{\theequation}{\arabic{section}.\arabic{equation}}
\nc{\sect}[1]{\section{#1}\setcounter{equation}{0}}
\nc{\sub}[1]{\subsection{#1}}
\nc{\subsub}[1]{\subsubsection{#1}}
\newcounter{subeq}
\nc{\beq}{\begin{equation}}
\nc{\beqa}{\begin{eqnarray}}
\nc{\eql}[1]{\label{Eqn#1}}
\nc{\eeq}{\end{equation}}
\nc{\eeqa}{\end{eqnarray}}
\nc{\noeqno}{\nonumber\\}
\nc{\eqref}[1]{(\ref{Eqn#1})}
\rnc{\aa}{\bar{a}}
\rnc{\AA}{\bar{A}}
\nc{\bb}{\bar{b}}
\nc{\B}[4]{B\!\!\left(\left.\!#2\,\begin{array}{c}#1\\#3\end{array}
\!\right|#4\right)}
\nc{\BB}[4]{\bar{B}\!\!\left(\left.\!#2\,\begin{array}{c}#1\\#3\end{array}
\!\right|#4\right)}
\nc{\BBB}{{\cal B}}
\rnc{\c}{\chi}
\nc{\cc}{\bar{c}}
\nc{\dd}{\bar{d}}
\nc{\ee}{\bar{e}}
\nc{\E}{{\cal E}}
\nc{\ep}{\epsilon}
\nc{\ev}{\mbox{\scriptsize even}}
\nc{\ff}{\bar{f}}
\rnc{\gg}{\bar{g}}
\nc{\G}{{\cal G}}
\nc{\GG}{\bar{\cal G}}
\nc{\I}[4]{I\!\left(\,\begin{array}{@{}cc@{}}#4&#3\\#1&#2\end{array}\,
\right)}
\rnc{\l}{\lambda}
\rnc{\ll}{\tilde{\lambda}}
\nc{\m}{\mu}
\nc{\mi}{\!-\!}
\nc{\mipl}{\!\mp\!}
\nc{\n}{\nu}
\nc{\N}{{\scriptscriptstyle N}}
\nc{\NN}{{\cal N}}
\nc{\odd}{\mbox{\scriptsize odd}}
\nc{\pa}{\pi_{\!a}}
\nc{\pl}{\!+\!}
\nc{\plmi}{\!\pm\!}
\rnc{\P}{{\cal P}}
\nc{\RRR}{{\cal R}}
\nc{\sm}[1]{{\scriptstyle #1}}
\nc{\smB}[4]{B\Big(\sm{#2}\begin{array}{c}\sm{#1}\\\sm{#3}\end{array}
\!\!\Big|#4\Big)}
\nc{\ssm}[1]{{\scriptscriptstyle #1}}
\nc{\SS}{\bar{S}}
\nc{\SSS}{{\cal S}}
\rnc{\t}{\vartheta_1}
\rnc{\tt}{\vartheta_2}
\nc{\ttt}{\vartheta_3}
\nc{\tttt}{\vartheta_4}
\nc{\th}{\vartheta}
\rnc{\vec}[1]{\mbox{\boldmath$#1$}}
\nc{\ww}{w_0}
\nc{\W}[5]{W\!\left(\,\begin{array}{@{}cc|}#4&#3\\#1&#2\end{array}\;#5\right)}
\nc{\WW}[5]{\overline{W}\!
\left(\,\begin{array}{@{}cc|}#4&#3\\#1&#2\end{array}\;#5\right)}
\nc{\x}{\xi}
\nc{\pos}[2]{\makebox(0,0)[#1]{$#2$}}
\nc{\spos}[2]{\makebox(0,0)[#1]{$\sm{#2}$}}
\nc{\text}[6]{\begin{picture}(#1,#2)
\put(#3,#4){\pos{#5}{\displaystyle#6}}\end{picture}}
\begin{document}
\begin{center}
\title{Solutions of the Boundary Yang-Baxter Equation for \ADE Models}
\author{Roger E. Behrend\footnote{E-mail: reb@maths.mu.oz.au} and
Paul A. Pearce\footnote{E-mail: pap@maths.mu.oz.au}}
\address{Department of Mathematics, University of Melbourne,\\Parkville,
Victoria 3052, Australia}
\begin{abstract}
\noindent  We present the general diagonal and, in some cases, non-diagonal
solutions of the boundary Yang-Baxter equation 
for a number of related interaction-round-a-face models, including the
standard and dilute $A_L$, $D_L$ and $E_{6,7,8}$ models.
\end{abstract}
\end{center}

\sect{Introduction}
A two-dimensional lattice spin model in statistical mechanics can be
considered as solvable with periodic boundary conditions if its bulk
Boltzmann weights satisfy the Yang-Baxter equation~\cite{Bax82b}, and
as additionally solvable with certain non-periodic boundary conditions if it
admits boundary weights which satisfy the boundary Yang-Baxter
equation~\cite{Skl88}.

Many such models are now known.  Restricting our attention to
interaction-round-a-face models, these are the $A_{\infty}$ or 
eight-vertex solid-on-solid model~\cite{FanHouShi95},
the $A^{(1)}_L$  or cyclic solid-on-solid models~\cite{ZhoBat96b},
the  $A_L$ or Andrews-Baxter-Forrester
models~\cite{AhnKoo96a,BehPeaObr96,BehPea96},
the dilute $A_L$ models~\cite{BatFriZho96},
and certain higher-rank models associated with $A_n^{(1)}$,
$B_n^{(1)}$, $C_n^{(1)}$, $D_n^{(1)}$ and
$A_n^{(2)}$~\cite{BatFriKunZho96}.  Here, we present general forms of
the boundary weights for some of these previously-considered models
and for some additional, related models.

We begin, in Section 2, by outlining the standard relations, including
the Yang-Baxter equation and the boundary Yang-Baxter equation, which
may be satisfied by the bulk and boundary weights of an
interaction-round-a-face model, and we define two important types of
boundary weight, diagonal and non-diagonal.  In Section 3, we consider
certain intertwining properties which may be satisfied by the bulk and
boundary weights of two appropriately-related interaction-round-a-face
models.  In Sections 4--9, we obtain boundary weights, mostly of the
diagonal type, which represent general solutions of the boundary
Yang-Baxter equation for various models.  These models, 
some of their relationships
and the sections in which they are considered are indicated in
Figure~1.  We conclude, in Section 10, with a discussion of 
general techniques for solving the boundary Yang-Baxter equation.
\setlength{\unitlength}{10mm}\begin{center}\begin{picture}(16.4,11.5)
\put(0,10.5){\framebox(2.7,1){\S4 $A_{\infty}$ Model}}
\put(4.8,10.5){\framebox(3.7,1){\S5 $A_L$ Models}}
\put(9.3,10.5){\framebox(3.7,1){\S6 $D_L$ Models}}
\put(0,8.5){\dashbox{0.1}(3.5,1){Critical $A_{\infty}$ Model}}
\put(4.8,8.5){\dashbox{0.1}(3.7,1){Critical $A_L$ Models}}
\put(9.3,8.5){\dashbox{0.1}(3.7,1){Critical $D_L$ Models}}
\put(13.8,8.5){\dashbox{0.1}(2.6,1){$E_{6,7,8}$ Models}}
\put(7,6.5){\framebox(7.2,1){\S7 Temperley-Lieb Models}}
\put(2.4,4){\framebox(5,1){\S8 Dilute $A_L$ Models}}
\put(2.4,2){\dashbox{0.1}(5,1){Critical Dilute $A_L$ Models}}
\put(8.2,2){\dashbox{0.1}(3.5,1){Dilute $D_L$ Models}}
\put(12.5,2){\dashbox{0.1}(3.9,1){Dilute $E_{6,7,8}$ Models}}
\put(5.4,0){\framebox(8,1){\S9 Dilute Temperley-Lieb Models}}
\put(3.5,9){\vector(1,0){1.3}}
\put(4.15,9.1){\makebox(0,0)[b]{\footnotesize rest'n}}
\put(2.7,11){\vector(1,0){2.1}}
\put(3.75,11.1){\makebox(0,0)[b]{\footnotesize restriction}}
\put(7.5,7.5){\vector(0,1){1}}
\put(7.45,8){\makebox(0,0)[r]{\footnotesize graph $=A_L$}}
\put(12,7.5){\vector(0,1){1}}
\put(11.95,8){\makebox(0,0)[r]{\footnotesize graph $=D_L$}}
\put(14,7.5){\vector(0,1){1}}
\put(14.1,8){\makebox(0,0)[l]{\footnotesize graph $=E_{6,7,8}$}}
\put(2.4,10.5){\vector(0,-1){1}}
\put(2.3,10){\makebox(0,0)[r]{\footnotesize criticality}}
\put(7.5,10.5){\vector(0,-1){1}}
\put(7.4,10){\makebox(0,0)[r]{\footnotesize criticality}}
\put(12,10.5){\vector(0,-1){1}}
\put(11.9,10){\makebox(0,0)[r]{\footnotesize criticality}}
\put(8.2,6.1){\vector(0,-1){0.75}}
\put(8.3,5.75){\makebox(0,0)[l]{\footnotesize dilution}}
\put(5.9,4){\vector(0,-1){1}}
\put(5.8,3.5){\makebox(0,0)[r]{\footnotesize criticality}}
\put(5.9,1){\vector(0,1){1}}
\put(5.85,1.5){\makebox(0,0)[r]{\footnotesize graph $=A_L$}}
\put(11,1){\vector(0,1){1}}
\put(10.95,1.5){\makebox(0,0)[r]{\footnotesize graph $=D_L$}}
\put(13,1){\vector(0,1){1}}
\put(13.1,1.5){\makebox(0,0)[l]{\footnotesize graph $=E_{6,7,8}$}}
\end{picture}\end{center}
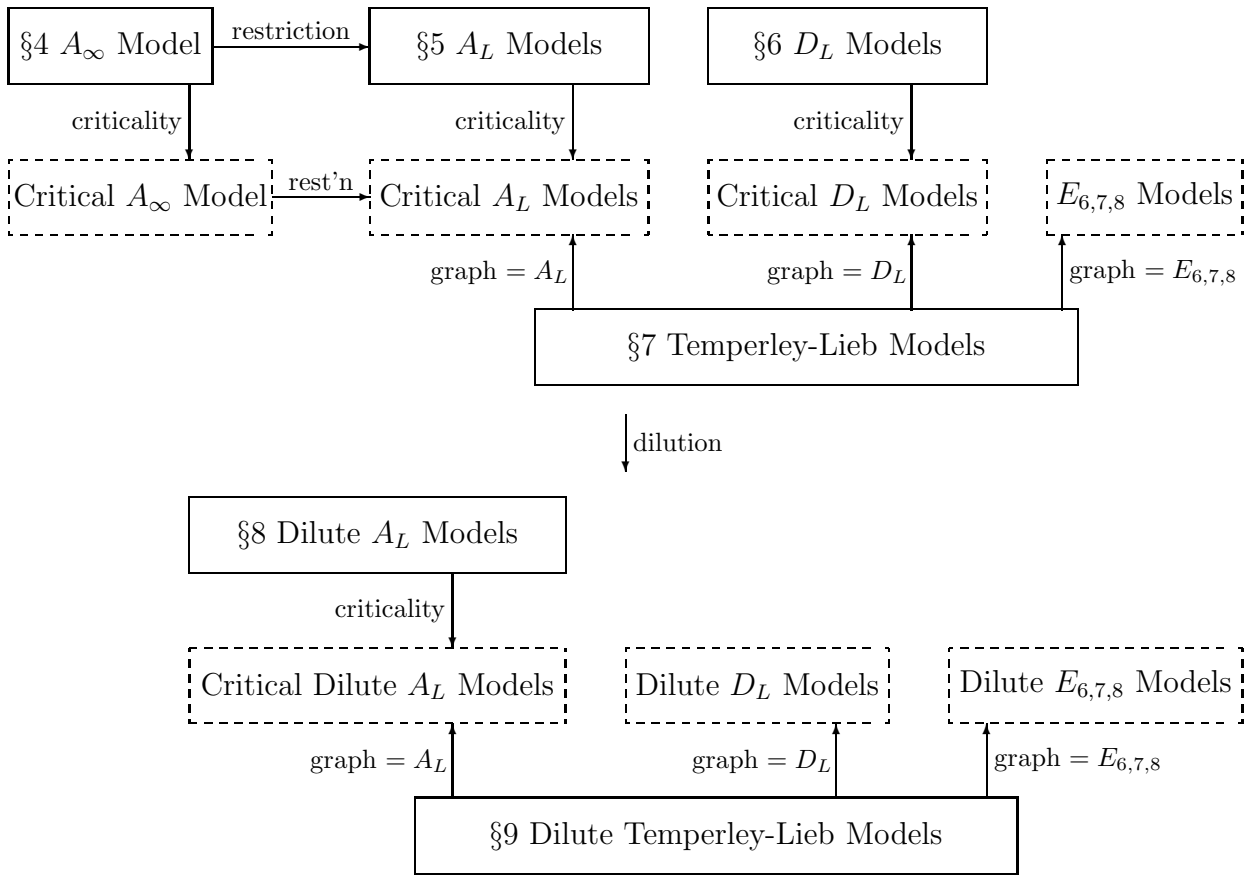
\begin{center}Figure 1. Models considered in this paper\end{center}

\sect{Interaction-Round-a-Face Models}
An interaction-round-a-face model is generally
associated with an adjacency graph $\G$ whose nodes 
correspond to the model's spin values and whose bonds
specify pairs of spin values which are allowed on neighbouring lattice sites.

The adjacency matrix $A$ associated with $\G$ is defined by
\[A_{ab}=\mbox{number of bonds of $\G$ which connect $a$ to $b$}\]
for any spin values $a$ and $b$.
Here, we consider cases in which $\G$ contains only bidirectional, single 
bonds, implying that $A$ is symmetric and that each of its entries is either
$0$ or $1$. However the formalism can be generalised straightforwardly to
accommodate directional or multiple bonds.

For interaction-round-a-face models with non-periodic boundary 
conditions, we associate a bulk weight $W$ with each set of spin values
$a$, $b$, $c$, $d$  satisfying $A_{ab}\:A_{cb}\:A_{dc}\:A_{da}=1$
and a boundary weight $B$ with each set of spin values 
$a$, $b$, $c$ satisfying $A_{ba}\:A_{bc}=1$~\cite{Kul96,BehPeaObr96}.
These weights are assumed to depend on a complex spectral parameter 
$u$ and are denoted
\setlength{\unitlength}{8mm}
\beq\eql{FW}\raisebox{-1\unitlength}[1\unitlength][1\unitlength]
{\text{3.2}{2}{1.6}{1}{}{\W{a}{b}{c}{d}{u}}
\text{1.4}{2}{0.7}{1}{}{=}
\begin{picture}(2,2)
\multiput(0.5,0.5)(1,0){2}{\line(0,1){1}}
\multiput(0.5,0.5)(0,1){2}{\line(1,0){1}}
\put(0.48,0.63){\spos{bl}{\searrow}}
\put(0.45,0.45){\spos{tr}{a}}\put(1.55,0.45){\spos{tl}{b}}
\put(1.55,1.55){\spos{bl}{c}}\put(0.45,1.55){\spos{br}{d}}
\put(1,1){\spos{}{u}}\end{picture}}\eeq
and
\setlength{\unitlength}{7mm}
\beq\eql{BFW}\raisebox{-1.5\unitlength}[1.5\unitlength][1\unitlength]
{\text{2.6}{3}{1.3}{1.5}{}{\B{a}{b}{c}{u}}
\text{1.6}{3}{0.8}{1.5}{}{=}
\begin{picture}(2,3)\put(1.5,0.5){\line(0,1){2}}
\put(0.5,1.5){\line(1,-1){1}}\put(0.5,1.5){\line(1,1){1}}
\put(1,1.5){\spos{l}{u}}
\put(1.55,2.55){\spos{bl}{a}}\put(0.4,1.5){\spos{r}{b}}
\put(1.55,0.45){\spos{tl}{c}}\put(2.2,1.4){\pos{b}{.}}\end{picture}}\eeq
Relations which may be satisfied by these weights are:\\
\prop{The Yang-Baxter equation,}
\beqa\lefteqn{\rule[-3.5ex]{0ex}{3.5ex}
\sum_{\stackrel{\scriptstyle g}{\scriptscriptstyle 
A_{fg}A_{gb}A_{gd}=1}}\!\!
\W{a}{b}{g}{f}{u\mi v}\;\W{b}{c}{d}{g}{u}\;
\W{g}{d}{e}{f}{v}\;=}\qquad\qquad\qquad\qquad\quad\noeqno
&&\sum_{\stackrel{\scriptstyle g}{\scriptscriptstyle
A_{ag}A_{eg}A_{gc}=1}}\!\!
\W{b}{c}{g}{a}{v}\;\W{a}{g}{e}{f}{u}\;\W{g}{c}{d}{e}{u\mi v}\eql{YBE}\eeqa
\setlength{\unitlength}{7mm}
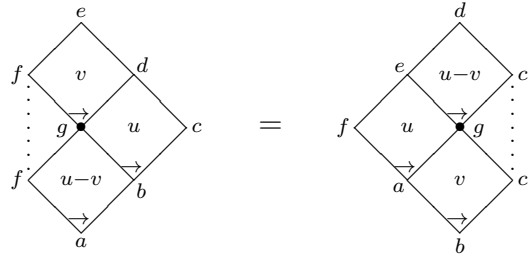
\begin{center}
\begin{picture}(4,5)\multiput(0.5,1.5)(1,-1){2}{\line(1,1){2}}
\multiput(0.5,3.5)(1,1){2}{\line(1,-1){2}}
\put(0.5,1.5){\line(1,-1){1}}\put(0.5,3.5){\line(1,1){1}}
\put(1.5,1.5){\spos{}{u-v}}\put(2.5,2.5){\spos{}{u}}
\put(1.5,3.5){\spos{}{v}}
\multiput(0.5,1.72)(0,0.26){7}{\spos{}{.}}
\multiput(1.45,0.66)(0,2){2}{\spos{b}{\rightarrow}}
\put(2.45,1.66){\spos{b}{\rightarrow}}
\put(1.5,0.4){\spos{t}{a}}\put(2.55,1.45){\spos{tl}{b}}
\put(3.6,2.5){\spos{l}{c}}\put(2.55,3.55){\spos{bl}{d}}
\put(1.5,4.6){\spos{b}{e}}\multiput(0.4,1.5)(0,2){2}{\spos{r}{f}}
\put(1.25,2.45){\spos{r}{g}}
\put(1.5,2.5){\spos{}{\bullet}}\end{picture}
\text{1.8}{5}{0.9}{2.5}{}{=}
\begin{picture}(4,5)\multiput(0.5,2.5)(1,1){2}{\line(1,-1){2}}
\multiput(0.5,2.5)(1,-1){2}{\line(1,1){2}}
\put(2.5,0.5){\line(1,1){1}}\put(2.5,4.5){\line(1,-1){1}}
\put(2.5,1.5){\spos{}{v}}\put(1.5,2.5){\spos{}{u}}
\put(2.5,3.5){\spos{}{u-v}}
\multiput(3.5,1.72)(0,0.26){7}{\spos{}{.}}
\multiput(2.45,0.66)(0,2){2}{\spos{b}{\rightarrow}}
\put(1.45,1.66){\spos{b}{\rightarrow}}
\put(1.45,1.45){\spos{tr}{a}}\put(2.5,0.4){\spos{t}{b}}
\multiput(3.6,1.5)(0,2){2}{\spos{l}{c}}
\put(2.5,4.6){\spos{b}{d}}\put(1.45,3.55){\spos{br}{e}}
\put(0.4,2.5){\spos{r}{f}}\put(2.75,2.45){\spos{l}{g}}
\put(2.5,2.5){\spos{}{\bullet}}\end{picture}\end{center}
for all $u$ and $v$ and all $a$, $b$, $c$, $d$, $e$, $f$ satisfying 
$A_{ab}\:A_{bc}\:A_{dc}\:A_{ed}\:A_{fe}\:A_{fa}=1$.

\prop{The boundary Yang-Baxter equation,}
\beqa\eql{BYBE}\lefteqn{\rule[-3.5ex]{0ex}{3.5ex}
\sum_{\stackrel{\scriptstyle
fg}{\scriptscriptstyle A_{cf}A_{fe}A_{fg}A_{bg}=1}}\!\!
\W{d}{e}{f}{c}{u\mi v}\;\B{g}{f}{e}{u}\;\W{f}{g}{b}{c}{u\pl v}\;
\B{a}{b}{g}{v}\;=}\qquad\qquad\\
&&\sum_{\stackrel{\scriptstyle
fg}{\scriptscriptstyle A_{cf}A_{fa}A_{fg}A_{dg}=1}}\!\!
\B{g}{d}{e}{v}\;\W{d}{g}{f}{c}{u\pl v}\;\B{a}{f}{g}{u}\;
\W{f}{a}{b}{c}{u\mi v}\nonumber\eeqa
\setlength{\unitlength}{7mm}\begin{center}
\begin{picture}(3,6)\multiput(0.5,1.5)(1,3){2}{\line(1,-1){1}}
\put(0.5,3.5){\line(1,-1){2}}\put(1.5,0.5){\line(1,1){1}}
\multiput(0.5,1.5)(0,2){2}{\line(1,1){2}}\put(2.5,1.5){\line(0,1){4}}
\multiput(0.5,1.72)(0,0.26){7}{\spos{}{.}}
\multiput(1.45,0.66)(0,2){2}{\spos{b}{\rightarrow}}
\put(1.5,1.5){\spos{}{u-v}}\put(1.5,3.5){\spos{}{u+v}}
\put(2,2.5){\spos{l}{u}}\put(2,4.5){\spos{l}{v}}
\put(2.55,5.55){\spos{bl}{a}}\put(1.45,4.55){\spos{br}{b}}
\multiput(0.4,1.5)(0,2){2}{\spos{r}{c}}\put(1.5,0.4){\spos{t}{d}}
\put(2.6,1.5){\spos{l}{e}}\put(1.27,2.45){\spos{r}{f}}
\put(2.68,3.5){\spos{l}{g}}
\multiput(1.5,2.5)(1,1){2}{\spos{}{\bullet}}\end{picture}
\text{1.8}{6}{0.9}{3}{}{=}
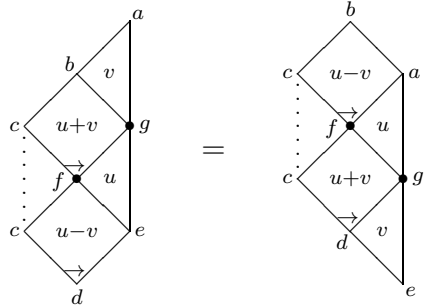
\begin{picture}(3,6)\multiput(0.5,2.5)(0,2){2}{\line(1,-1){2}}
\put(1.5,5.5){\line(1,-1){1}}\multiput(1.5,1.5)(-1,3){2}{\line(1,1){1}}
\put(0.5,2.5){\line(1,1){2}}\put(2.5,0.5){\line(0,1){4}}
\multiput(0.5,2.72)(0,0.26){7}{\spos{}{.}}
\multiput(1.45,1.66)(0,2){2}{\spos{b}{\rightarrow}}
\put(1.5,2.5){\spos{}{u+v}}\put(1.5,4.5){\spos{}{u-v}}
\put(2,1.5){\spos{l}{v}}\put(2,3.5){\spos{l}{u}}
\put(2.6,4.5){\spos{l}{a}}\put(1.5,5.6){\spos{b}{b}}
\multiput(0.4,2.5)(0,2){2}{\spos{r}{c}}\put(1.45,1.45){\spos{tr}{d}}
\put(2.55,0.45){\spos{tl}{e}}\put(1.27,3.45){\spos{r}{f}}
\put(2.68,2.5){\spos{l}{g}}
\multiput(2.5,2.5)(-1,1){2}{\spos{}{\bullet}}\end{picture}
\end{center}
for all $u$ and $v$ and all $a$, $b$, $c$, $d$, $e$ satisfying 
$A_{ba}\:A_{cb}\:A_{cd}\:A_{de}=1$.

\prop{The initial condition,}
\beq\eql{IC}\W{a}{b}{c}{d}{0}\;=\;\delta_{ac}\eeq
for all $a$, $b$, $c$, $d$ satisfying 
$A_{ab}\:A_{cb}\:A_{dc}\:A_{da}=1$.

\prop{The boundary initial condition,}
\beq\eql{BIC}\B{a}{b}{c}{0}\;=\;\chi_a\:\delta_{ac}\eeq
for all $a$, $b$, $c$ satisfying 
$A_{ba}\:A_{bc}=1$, where $\chi_a$ are fixed, model-dependent factors.

\prop{Invariance of the bulk weights under a symmetry transformation
$Z$ of the graph $\G$,}
\beq\eql{ZS}\W{a}{b}{c}{d}{u}\;=\;\W{Z(a)}{Z(b)}{Z(c)}{Z(d)}{u}\eeq
for all $u$ and all $a$, $b$, $c$, $d$ satisfying 
$A_{ab}\:A_{cb}\:A_{dc}\:A_{da}=1$.

\prop{Invariance of the boundary weights under a symmetry
transformation $Z$ of the graph $\G$,}
\beq\eql{BZS}\B{a}{b}{c}{u}\;=\;\B{Z(a)}{Z(b)}{Z(c)}{u}\eeq
for all $u$ and all $a$, $b$, $c$ satisfying 
$A_{ba}\:A_{bc}=1$.

\prop{Reflection symmetry of the bulk weights,}
\beqa\eql{RS1}\rule[-3.5ex]{0ex}{3.5ex}\W{a}{b}{c}{d}{u}\;=\;
\W{a}{d}{c}{b}{u}\;=\;\W{c}{b}{a}{d}{u}\eeqa 
for all $u$ and all $a$, $b$, $c$, $d$ satisfying 
$A_{ab}\:A_{cb}\:A_{dc}\:A_{da}=1$.

\prop{Reflection symmetry of the boundary weights,}
\beq\eql{BRS}\B{a}{b}{c}{u}\;=\;\B{c}{b}{a}{u}\eeq
for all $u$ and all $a$, $b$, $c$ satisfying 
$A_{ba}\:A_{bc}=1$.

\prop{Crossing symmetry,}
\beq\eql{CS}\W{a}{b}{c}{d}{u}\;=\;
\left(\frac{S_a\,S_c}{S_b\,S_d}\right)^{\!\!\ssm{1\!/2}}\;
\W{d}{c}{b}{a}{\l\mi u}\eeq
for all $u$ and all $a$, $b$, $c$, $d$ satisfying 
$A_{ab}\:A_{cb}\:A_{dc}\:A_{da}=1$, where the crossing 
parameter $\l$ and crossing factors $S_a$ are fixed for a 
particular model.

\prop{Boundary crossing symmetry,}
\beq\eql{BCS}\eta_a(u)\;\B{a}{b}{c}{u}\;=\;
\sum_{\stackrel{\scriptstyle d}{\scriptscriptstyle A_{da}A_{dc}=1}}
\left(\frac{S_d^2}{S_a\,S_c}\right)^{\!\!\ssm{1\!/2}}
\;\W{c}{d}{a}{b}{2u}\;\B{a}{d}{c}{\l\mi u}\eeq
\setlength{\unitlength}{7mm}\begin{center}
\text{1.2}{3}{0.6}{1.5}{}{\eta_a(u)}
\begin{picture}(2,3)\put(1.5,0.5){\line(0,1){2}}
\put(0.5,1.5){\line(1,-1){1}}\put(0.5,1.5){\line(1,1){1}}
\put(1,1.5){\spos{l}{u}}
\put(1.55,2.55){\spos{bl}{a}}\put(0.4,1.5){\spos{r}{b}}
\put(1.55,0.45){\spos{tl}{c}}\end{picture}
\text{1.6}{3}{0.8}{1.5}{}{=}
\text{2.4}{3}{1.2}{1.5}{}{\left(\frac{S_d^2}{S_a\,S_c}
\right)^{\!\!\ssm{1\!/2}}}
\begin{picture}(4,3)\put(0.5,1.5){\line(1,-1){1}}
\put(0.5,1.5){\line(1,1){1}}\put(1.5,0.5){\line(1,1){2}}
\put(1.5,2.5){\line(1,-1){2}}\put(3.5,0.5){\line(0,1){2}}
\put(1.45,0.66){\spos{b}{\rightarrow}}
\put(1.5,1.5){\spos{}{2u}}\put(2.75,1.5){\spos{l}{\l\mi u}}
\put(1.5,2.6){\spos{b}{a}}\put(3.55,2.55){\spos{bl}{a}}
\put(0.4,1.5){\spos{r}{b}}\put(1.5,0.4){\spos{t}{c}}
\put(3.55,0.45){\spos{tl}{c}}\put(2.5,1.3){\spos{t}{d}}
\multiput(1.72,0.5)(0.26,0){7}{\spos{}{.}}
\multiput(1.72,2.5)(0.26,0){7}{\spos{}{.}}
\put(2.5,1.5){\spos{}{\bullet}}\end{picture}\end{center}
for all $u$ and all $a$, $b$, $c$ satisfying 
$A_{ba}\:A_{bc}=1$, where $\l$ and $S_a$ are the same as 
in~\eqref{CS} and $\eta_a$ are fixed, model-dependent functions.

\prop{The inversion relation,}
\beq\eql{IR}
\sum_{\stackrel{\scriptstyle e}{\scriptscriptstyle A_{de}A_{eb}=1}}\!\!
\W{a}{b}{e}{d}{u}\;\W{e}{b}{c}{d}{\!\mi u}\;=\;
\rho(u)\,\rho(-u)\:\delta_{ac}\eeq
\setlength{\unitlength}{7mm}\begin{center}
\begin{picture}(3,5)
\multiput(1.5,0.5)(-1,3){2}{\line(1,1){1}}
\multiput(0.5,1.5)(1,3){2}{\line(1,-1){1}}
\put(0.5,1.5){\line(1,1){2}}\put(0.5,3.5){\line(1,-1){2}}
\multiput(0.5,1.72)(0,0.26){7}{\spos{}{.}}
\multiput(2.5,1.72)(0,0.26){7}{\spos{}{.}}
\multiput(1.45,0.66)(0,2){2}{\spos{b}{\rightarrow}}
\put(1.5,1.5){\spos{}{u}}\put(1.35,3.5){\spos{}{-u}}
\put(1.5,0.4){\spos{t}{a}}\multiput(2.6,1.5)(0,2){2}{\spos{l}{b}}
\put(1.5,4.6){\spos{b}{c}}\multiput(0.4,1.5)(0,2){2}{\spos{r}{d}}
\put(1.25,2.47){\spos{r}{e}}
\put(1.5,2.5){\spos{}{\bullet}}\end{picture}
\text{1.8}{5}{0.9}{2.5}{}{=}
\text{3.4}{5}{1.7}{2.5}{}{\rho(u)\,\rho(-u)\:\delta_{ac}}\end{center}
for all $u$ and all $a$, $b$, $c$, $d$ 
satisfying $A_{ab}\:A_{cb}\:A_{dc}\:A_{da}=1$,
where $\rho$ is a fixed, model-dependent function.

\prop{The boundary inversion relation,}
\beq\eql{BIR}
\sum_{\stackrel{\scriptstyle d}{\scriptscriptstyle A_{bd}=1}}
\:\B{d}{b}{c}{u}\;\B{a}{b}{d}{\!\mi u}\;=\;\hat{\rho}_a(u)\:\delta_{ac}\eeq
\setlength{\unitlength}{7mm}\begin{center}
\begin{picture}(2,5)\multiput(0.5,1.5)(0,2){2}{\line(1,-1){1}}
\multiput(0.5,1.5)(0,2){2}{\line(1,1){1}}\put(1.5,0.5){\line(0,1){4}}
\multiput(0.5,1.72)(0,0.26){7}{\spos{}{.}}
\put(1,1.5){\spos{l}{u}}\put(0.7,3.5){\spos{l}{-u}}
\put(1.55,4.55){\spos{bl}{a}}\multiput(0.4,1.5)(0,2){2}{\spos{r}{b}}
\put(1.55,0.45){\spos{tl}{c}}\put(1.68,2.5){\spos{l}{d}}
\put(1.5,2.5){\spos{}{\bullet}}\end{picture}
\text{1.8}{5}{0.9}{2.5}{}{=}
\text{2.2}{5}{1.1}{2.5}{}{\hat{\rho}_a(u)\:\delta_{ac}}
\end{center}
for all $u$ and all $a$, $b$, $c$ satisfying $A_{ba}\:A_{bc}=1$,
where $\hat{\rho}_a$ are fixed, model-dependent functions.

These relations are not all independent.  For example, it can be shown that
the boundary inversion relation is a consequence of the
initial condition and the boundary Yang-Baxter equation,
and that boundary crossing symmetry is a consequence of the
initial condition, bulk crossing symmetry and the boundary Yang-Baxter 
equation. 

We note that the boundary Yang-Baxter equation used here differs from the
left and right reflection equations introduced in~\cite{BehPeaObr96}.  
However if crossing symmetry is satisfied then the equations are
effectively equivalent.

The bulk and boundary weights can be used to define matrices whose rows
and columns are labelled by elements of 
$\{(a_0,\ldots,a_{\N\pl1})\,|\,
A_{a_0a_1}\:A_{a_1a_2}\,\ldots\,A_{a_Na_{N\pl1}}=1\}$, the set of paths
of length $N\pl1$.
The entries of the $j$th bulk face transfer matrix, for $1\leq j\leq N$,
are defined by
\beq\eql{FTM}X_j(u)_{(a_0\ldots a_{\N\pl1}),(b_0\ldots b_{\N\pl1})}
\;=\;\prod_{k=0}^{j\mi1}\delta_{a_kb_k}\;
\W{a_j}{a_{j\pl1}}{b_j}{a_{j\mi1}}{u}\;
\prod_{k=j\pl1}^{\N\pl1}\delta_{a_kb_k}\eeq
and the entries of the boundary face transfer matrix are defined by
\beq\eql{BFTM}K(u)_{(a_0\ldots a_{\N\pl1}),(b_0\ldots b_{\N\pl1})}
\;=\;\prod_{k=0}^{\N}\delta_{a_kb_k}\;
\B{b_{\N\pl1}}{a_\N}{a_{\N\pl1}}{u}\;.\eeq
In terms of these matrices, the Yang-Baxter equation is
\beq\eql{MYBE}X_j(u\mi v)\;X_{j\pl1}(u)\;X_j(v)\;=\;
X_{j\pl1}(v)\;X_j(u)\;X_{j\pl1}(u\mi v)\eeq
and the boundary Yang-Baxter equation is
\beq\eql{MBYBE}X_\N(u\mi v)\;K(u)\;X_\N(u\pl v)\;K(v)\;=\;
K(v)\;X_\N(u\pl v)\;K(u)\;X_\N(u\mi v)\;.\eeq
In the study of exactly solvable interaction-round-a-face models with 
non-periodic boundary conditions, we generally begin with bulk weights which
satisfy the Yang-Baxter equation and then attempt to solve the boundary
Yang-Baxter equation for corresponding boundary weights.
Two classes of solution which are of particular interest, since they are 
needed for fixed and free boundary conditions respectively, are diagonal
solutions, for which
\beq\B{a}{b}{c}{u}=0\quad\mbox{whenever $a\neq c$}\eeq
and non-diagonal solutions, for which
\beq\B{a}{b}{c}{u}\neq0\quad\mbox{ for all $a$, $b$, $c$}\;.\eeq
We note that if a diagonal solution satisfies the boundary initial condition,
then $\chi$ is given by
\beq\eql{DBIC}\chi_a\;=\;\B{a}{b}{a}{0}\eeq
for any $b$ satisfying $A_{ab}=1$,
and if a diagonal solution satisfies the boundary inversion relation,
then $\hat{\rho}$ is given by
\beq\eql{DBIR}\hat{\rho}_a(u)\;=\;\B{a}{b}{a}{\mi u}\:\B{a}{b}{a}{u}\eeq
for any $b$ satisfying $A_{ab}=1$.

\sect{Intertwiners}
In this section we outline the formalism of 
intertwiners~\cite{DifZub90,Roc90,Dif92,PeaZho93}.
We are considering two interaction-round-a-face models, 
with respective finite 
adjacency graphs $\G$ and $\GG$, adjacency matrices $A$ and $\AA$, spin
values $a,b,\ldots$ and $\aa,\bb,\ldots$, crossing factors $S_a$ and 
$\SS_{\aa}$, bulk weights $W$ and $\overline{W}$, and boundary 
weights $B$ and $\bar{B}$.

We now assume that there exists an intertwiner graph $\G'$ each of whose
bonds connects a node of $\G$ to a node of $\GG$, and whose
adjacency matrix $C$, defined by
\[C_{\!a\aa}=\mbox{number of bonds of $\G'$ which connect $a$ to $\aa$}\]
for any spin values $a$ and $\aa$, satisfies the intertwining relation
\beq\eql{I}AC=C\AA\;.\eeq
Here, we consider cases in which $\G'$ contains only single bonds,
so that each entry of $C$ is either $0$ or $1$.

We now associate an interwiner cell $I$ with 
each set of spin values $a$, $b$, $\aa$, $\bb$ satisfying 
$A_{ba}\:\AA_{\bb\aa}\:C_{\!a\aa}\:C_{\!b\bb}=1$, 
\setlength{\unitlength}{8mm}
\beq\raisebox{-1\unitlength}[1\unitlength][1\unitlength]
{\text{2.4}{2}{1.2}{1}{}{\I{a}{\aa}{\bb}{b}}
\text{1.4}{2}{0.7}{1}{}{=}
\begin{picture}(2,2)
\put(0.5,0.5){\line(0,1){1}}
\multiput(0.5,0.5)(0.15,0){7}{\line(1,0){0.1}}
\multiput(0.5,1.5)(0.15,0){7}{\line(1,0){0.1}}
\thicklines\put(1.5,0.5){\line(0,1){1}}
\put(0.48,0.63){\spos{bl}{\searrow}}
\put(0.45,0.45){\spos{tr}{a}}\put(1.55,0.45){\spos{tl}{\aa}}
\put(1.55,1.55){\spos{bl}{\bb}}\put(0.45,1.55){\spos{br}{b}}
\put(2.2,0.9){\pos{b}{.}}
\end{picture}}\eeq
Relations which may be satisfied by the intertwiner cells are:\\
\prop{The first intertwiner inversion relation,}
\setlength{\unitlength}{8mm}
\beq\eql{FIIR}\raisebox{-1.5\unitlength}[1.5\unitlength][1.5\unitlength]
{\text{6.4}{3}{3.2}{1.3}{}
{\sum_{\stackrel{\scriptstyle b}{\scriptscriptstyle A_{ab}C_{\!b\bb}=1}}\!\!
\I{b}{\bb}{\aa}{a}\;\I{b}{\bb}{\cc}{a}}
\text{1.4}{3}{0.7}{1.5}{}{=}
\begin{picture}(3,3)
\put(0.5,1.5){\line(1,0){1}}\put(1.5,1.5){\line(0,1){1}}
\multiput(0.5,0.5)(0,0.15){7}{\line(0,1){0.1}}
\multiput(1.5,0.5)(0,0.15){7}{\line(0,1){0.1}}
\multiput(1.5,1.5)(0.15,0){7}{\line(1,0){0.1}}
\multiput(1.5,2.5)(0.15,0){7}{\line(1,0){0.1}}
\thicklines\put(0.5,0.5){\line(1,0){1}}\put(2.5,1.5){\line(0,1){1}}
\multiput(0.61,1.61)(0.13,0.13){7}{\spos{}{.}}
\multiput(1.61,0.61)(0.13,0.13){7}{\spos{}{.}}
\put(1.5,1.5){\spos{}{\bullet}}
\put(1.48,1.63){\spos{bl}{\searrow}}
\put(1.44,1.49){\spos{tr}{\searrow}}
\put(0.45,0.45){\spos{tr}{\aa}}
\multiput(1.55,0.45)(1,1){2}{\spos{tl}{\bb}}
\put(2.55,2.55){\spos{bl}{\cc}}
\multiput(0.45,1.55)(1,1){2}{\spos{br}{a}}
\put(1.6,1.4){\spos{tl}{b}}\end{picture}
\text{1.4}{3}{0.7}{1.5}{}{=}
\text{1}{3}{0.5}{1.5}{}{\delta_{\aa\cc}}}\eeq
for all $a$, $\aa$, $\bb$, $\cc$ 
satisfying $\AA_{\aa\bb}\:\AA_{\cc\bb}\:C_{\!a\aa}\:C_{\!a\cc}=1$.

\prop{The second intertwiner inversion relation,}
\setlength{\unitlength}{8mm}
\beq\eql{SIIR}\raisebox{-1.5\unitlength}[1.5\unitlength][1.5\unitlength]
{\text{7.6}{3}{3.8}{1.3}{}
{\sum_{\stackrel{\scriptstyle b}{\scriptscriptstyle A_{ba}C_{\!b\bb}=1}}
\frac{\SS_{\aa}\,S_b}{\SS_{\bb}\,S_a}\;\,
\I{a}{\cc}{\bb}{b}\;\I{a}{\aa}{\bb}{b}}
\text{1.4}{3}{0.7}{1.5}{}{=}
\text{1.4}{3}{0.7}{1.5}{}{\frac{\SS_{\aa}\,S_b}{\SS_{\bb}\,S_a}}
\begin{picture}(3,3)
\put(0.5,1.5){\line(1,0){1}}\put(1.5,1.5){\line(0,-1){1}}
\multiput(0.5,1.5)(0,0.15){7}{\line(0,1){0.1}}
\multiput(1.5,1.5)(0,0.15){7}{\line(0,1){0.1}}
\multiput(1.5,0.5)(0.15,0){7}{\line(1,0){0.1}}
\multiput(1.5,1.5)(0.15,0){7}{\line(1,0){0.1}}
\thicklines\put(0.5,2.5){\line(1,0){1}}\put(2.5,0.5){\line(0,1){1}}
\multiput(0.61,1.39)(0.13,-0.13){7}{\spos{}{.}}
\multiput(1.61,2.39)(0.13,-0.13){7}{\spos{}{.}}
\put(1.5,1.5){\spos{}{\bullet}}
\put(0.56,1.58){\spos{bl}{\nwarrow}}
\put(1.48,0.63){\spos{bl}{\searrow}}
\put(0.45,2.55){\spos{br}{\cc}}
\multiput(0.45,1.45)(1,-1){2}{\spos{tr}{a}}
\put(2.55,0.45){\spos{tl}{\aa}}
\multiput(1.55,2.55)(1,-1){2}{\spos{bl}{\bb}}
\put(1.4,1.4){\spos{tr}{b}}\end{picture}
\text{1.4}{3}{0.7}{1.5}{}{=}
\text{1}{3}{0.5}{1.5}{}{\delta_{\aa\cc}}}\eeq
for all $a$, $\aa$, $\bb$, $\cc$ 
satisfying $\AA_{\bb\aa}\:\AA_{\bb\cc}\:C_{\!a\aa}\:C_{\!a\cc}=1$.

\prop{The bulk weight intertwining relation,}
\beqa\lefteqn{\rule[-3.5ex]{0ex}{3.5ex}
\sum_{\stackrel{\scriptstyle d}{\scriptscriptstyle
A_{ad}A_{dc}C_{\!d\bb}=1}}\!\!
\W{b}{c}{d}{a}{u}\;\I{d}{\bb}{\cc}{a}\;\I{c}{\aa}{\bb}{d}
\;=}\qquad\qquad\qquad\qquad\qquad\qquad\noeqno
&&\sum_{\stackrel{\scriptstyle \dd}{\scriptscriptstyle
\AA_{\cc\dd}\AA_{\dd\aa}C_{\!b\dd}=1}}\!\!
\I{b}{\dd}{\cc}{a}\;\I{c}{\aa}{\dd}{b}\;\WW{\dd}{\aa}{\bb}{\cc}{u}
\eql{FWIR}\eeqa
\setlength{\unitlength}{7mm}\begin{center}
\begin{picture}(3,4)
\multiput(0.5,1.5)(1,1){2}{\line(1,-1){1}}
\multiput(0.5,1.5)(1,-1){2}{\line(1,1){1}}
\put(0.5,2.5){\line(1,0){2}}
\multiput(0.5,2.5)(0,0.15){7}{\line(0,1){0.1}}
\multiput(1.5,2.5)(0,0.15){7}{\line(0,1){0.1}}
\multiput(2.5,2.5)(0,0.15){7}{\line(0,1){0.1}}
\thicklines\put(0.5,3.5){\line(1,0){2}}
\multiput(0.5,1.74)(0,0.26){3}{\spos{}{.}}
\multiput(2.5,1.74)(0,0.26){3}{\spos{}{.}}
\put(1.5,2.5){\spos{}{\bullet}}
\multiput(1.44,2.6)(1,0){2}{\spos{br}{\nearrow}}
\put(1.45,0.66){\spos{b}{\rightarrow}}
\put(1.5,1.5){\spos{}{u}}\multiput(0.4,1.5)(0,1){2}{\spos{r}{a}}
\put(1.5,0.4){\spos{t}{b}}\multiput(2.6,1.5)(0,1){2}{\spos{l}{c}}
\put(1,2.44){\spos{t}{d}}\put(2.55,3.55){\spos{bl}{\aa}}
\put(1.5,3.6){\spos{b}{\bb}}\put(0.45,3.55){\spos{br}{\cc}}
\end{picture}
\text{1.8}{4}{0.9}{2}{}{=}
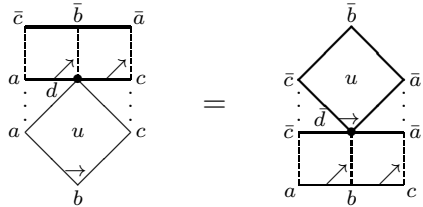
\begin{picture}(3,4)
\put(0.5,0.5){\line(1,0){2}}
\multiput(0.5,0.5)(0,0.15){7}{\line(0,1){0.1}}
\multiput(1.5,0.5)(0,0.15){7}{\line(0,1){0.1}}
\multiput(2.5,0.5)(0,0.15){7}{\line(0,1){0.1}}
\thicklines\multiput(0.5,2.5)(1,1){2}{\line(1,-1){1}}
\multiput(0.5,2.5)(1,-1){2}{\line(1,1){1}}
\put(0.5,1.5){\line(1,0){2}}
\multiput(0.5,1.74)(0,0.26){3}{\spos{}{.}}
\multiput(2.5,1.74)(0,0.26){3}{\spos{}{.}}
\put(1.5,1.5){\spos{}{\bullet}}
\multiput(1.44,0.6)(1,0){2}{\spos{br}{\nearrow}}
\put(1.45,1.66){\spos{b}{\rightarrow}}
\put(1.5,2.5){\spos{}{u}}\put(0.45,0.45){\spos{tr}{a}}
\put(1.5,0.4){\spos{t}{b}}\put(2.55,0.45){\spos{tl}{c}}
\multiput(2.6,1.5)(0,1){2}{\spos{l}{\aa}}
\put(1.5,3.6){\spos{b}{\bb}}\multiput(0.4,1.5)(0,1){2}{\spos{r}{\cc}}
\put(0.9,1.61){\spos{b}{\dd}}
\end{picture}
\end{center}
for all $u$ and all $a$, $b$, $c$, $\aa$, $\bb$, $\cc$ 
satisfying $A_{ab}\:A_{bc}\:\AA_{\cc\bb}\:\AA_{\bb\aa}
\:C_{\!a\cc}\:C_{\!c\aa}=1$.

\prop{The boundary weight intertwining relation,}
\beq\eql{BWIR}
\sum_{\stackrel{\scriptstyle c}{\scriptscriptstyle A_{ac}C_{\!c\aa}=1}}\!\!
\B{c}{a}{d}{u}\;\I{c}{\aa}{\bb}{a}\;=\;
\sum_{\stackrel{\scriptstyle\cc}{\scriptscriptstyle 
\AA_{\bb\cc}C_{\!b\cc}=1}}\!\!
\I{b}{\cc}{\bb}{a}\;\BB{\aa}{\bb}{\cc}{u}\eeq
\setlength{\unitlength}{7mm}
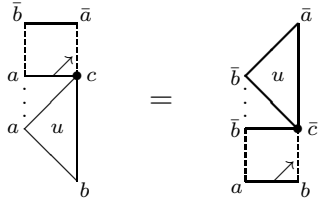
\begin{center}
\begin{picture}(2,4)
\put(0.5,1.5){\line(1,-1){1}}\put(0.5,1.5){\line(1,1){1}}
\put(1.5,0.5){\line(0,1){2}}\put(0.5,2.5){\line(1,0){1}}
\multiput(0.5,2.5)(0,0.15){7}{\line(0,1){0.1}}
\multiput(1.5,2.5)(0,0.15){7}{\line(0,1){0.1}}
\thicklines\put(0.5,3.5){\line(1,0){1}}
\multiput(0.5,1.74)(0,0.26){3}{\spos{}{.}}
\put(1.5,2.5){\spos{}{\bullet}}
\put(1.44,2.6){\spos{br}{\nearrow}}
\put(1,1.5){\spos{l}{u}}
\put(1.55,3.55){\spos{bl}{\aa}}\put(0.45,3.55){\spos{br}{\bb}}
\multiput(0.4,1.5)(0,1){2}{\spos{r}{a}}
\put(1.55,0.48){\spos{tl}{b}}\put(1.68,2.5){\spos{l}{c}}
\end{picture}
\text{1.8}{4}{0.9}{2}{}{=}
\begin{picture}(3,4)
\put(0.5,0.5){\line(1,0){1}}
\multiput(0.5,0.5)(0,0.15){7}{\line(0,1){0.1}}
\multiput(1.5,0.5)(0,0.15){7}{\line(0,1){0.1}}
\thicklines
\put(0.5,2.5){\line(1,-1){1}}\put(0.5,2.5){\line(1,1){1}}
\put(1.5,1.5){\line(0,1){2}}\put(0.5,1.5){\line(1,0){1}}
\multiput(0.5,1.74)(0,0.26){3}{\spos{}{.}}
\put(1.5,1.5){\spos{}{\bullet}}
\put(1.44,0.6){\spos{br}{\nearrow}}
\put(1,2.5){\spos{l}{u}}
\put(1.55,3.55){\spos{bl}{\aa}}\multiput(0.4,1.5)(0,1){2}{\spos{r}{\bb}}
\put(0.45,0.45){\spos{tr}{a}}\put(1.55,0.48){\spos{tl}{b}}
\put(1.68,1.5){\spos{l}{\cc}}
\end{picture}\end{center}
for all $u$ and all $a$, $b$, $\aa$, $\bb$
satisfying $A_{ab}\:\AA_{\bb\aa}\:C_{\!a\bb}=1$.

The bulk and boundary weight intertwining relations can be combined
with the intertwiner inversion relations to give expressions for the
weights of one model in terms of those of the other model.
For example, assuming the first intertwiner inversion relation,
we find that the bulk weight intertwining relation is equivalent to
\beqa\eql{FWI1}
\lefteqn{\WW{\aa}{\bb}{\cc}{\dd}{u}\;\delta_{\dd\dd'}\;=}\qquad\\
&&\sum_{\stackrel{\scriptstyle abc}{\scriptscriptstyle 
A_{ab}A_{cb}A_{dc}A_{da}C_{\!a\aa}C_{\!b\bb}C_{\!c\cc}=1}}\!\!
\I{a}{\aa}{\dd}{d}\;\I{b}{\bb}{\aa}{a}\;\I{b}{\bb}{\cc}{c}\;
\I{c}{\cc}{\dd'}{d}\;\W{a}{b}{c}{d}{u}\nonumber\eeqa
\setlength{\unitlength}{7mm}
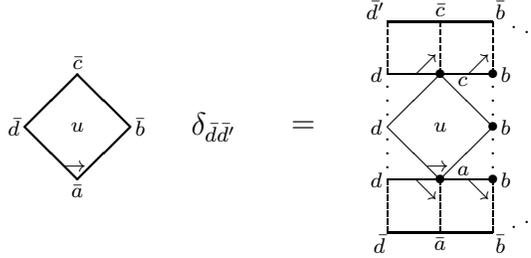
\begin{center}
\begin{picture}(3,5)
\thicklines\multiput(0.5,2.5)(1,1){2}{\line(1,-1){1}}
\multiput(0.5,2.5)(1,-1){2}{\line(1,1){1}}
\put(1.45,1.66){\spos{b}{\rightarrow}}
\put(1.5,2.5){\spos{}{u}}
\put(1.5,1.4){\spos{t}{\aa}}\put(2.6,2.5){\spos{l}{\bb}}
\put(1.5,3.6){\spos{b}{\cc}}\put(0.4,2.5){\spos{r}{\dd}}
\end{picture}
\text{1.5}{5}{0.9}{2.5}{}{\delta_{\dd\dd'}}
\text{1.8}{5}{0.9}{2.5}{}{=}
\begin{picture}(3.6,5)
\multiput(0.5,2.5)(1,1){2}{\line(1,-1){1}}
\multiput(0.5,2.5)(1,-1){2}{\line(1,1){1}}
\multiput(0.5,1.5)(0,2){2}{\line(1,0){2}}
\multiput(0.5,0.5)(0,0.15){7}{\line(0,1){0.1}}
\multiput(1.5,0.5)(0,0.15){7}{\line(0,1){0.1}}
\multiput(2.5,0.5)(0,0.15){7}{\line(0,1){0.1}}
\multiput(0.5,3.5)(0,0.15){7}{\line(0,1){0.1}}
\multiput(1.5,3.5)(0,0.15){7}{\line(0,1){0.1}}
\multiput(2.5,3.5)(0,0.15){7}{\line(0,1){0.1}}
\thicklines\multiput(0.5,0.5)(0,4){2}{\line(1,0){2}}
\multiput(0.5,1.74)(0,0.26){3}{\spos{}{.}}
\multiput(2.5,1.74)(0,0.26){3}{\spos{}{.}}
\multiput(0.5,2.74)(0,0.26){3}{\spos{}{.}}
\multiput(2.5,2.74)(0,0.26){3}{\spos{}{.}}
\multiput(3.3,0.94)(0,0.26){13}{\spos{}{.}}
\put(2.9,0.6){\spos{}{.}}\put(3.15,0.7){\spos{}{.}}
\put(2.9,4.4){\spos{}{.}}\put(3.15,4.3){\spos{}{.}}
\multiput(1.5,1.5)(0,2){2}{\spos{}{\bullet}}
\multiput(2.5,1.5)(0,1){3}{\spos{}{\bullet}}
\multiput(1.44,3.6)(1,0){2}{\spos{br}{\nearrow}}
\multiput(1.44,1.49)(1,0){2}{\spos{tr}{\searrow}}
\put(1.45,1.66){\spos{b}{\rightarrow}}
\put(1.5,2.5){\spos{}{u}}
\put(0.45,0.45){\spos{tr}{\dd}}\put(1.5,0.4){\spos{t}{\aa}}
\put(2.55,0.45){\spos{tl}{\bb}}\multiput(2.65,1.5)(0,1){3}{\spos{l}{b}}
\put(2.55,4.55){\spos{bl}{\bb}}\put(1.5,4.6){\spos{b}{\cc}}
\put(0.45,4.55){\spos{br}{\dd'}}\multiput(0.4,1.5)(0,1){3}{\spos{r}{d}}
\put(1.94,1.57){\spos{b}{a}}\put(1.92,3.45){\spos{t}{c}}
\end{picture}\end{center}
for all $u$ and all $d$, $\aa$, $\bb$, $\cc$, $\dd$, $\dd'$ satisfying
$\AA_{\aa\bb}\:\AA_{\cc\bb}\:\AA_{\dd'\cc}\:\AA_{\dd\aa}
\:C_{\!d\dd}\:C_{\!d\dd'}=1$. Similarly,
assuming the second intertwiner inversion relation,
we find that the bulk weight intertwining relation is equivalent to
\beqa\eql{FWI2}
\lefteqn{\WW{\aa}{\bb}{\cc}{\dd}{u}\;\delta_{\bb\bb'}\;=}\\
&&\sum_{\stackrel{\scriptstyle acd}{\scriptscriptstyle 
A_{ab}A_{cb}A_{dc}A_{da}C_{\!a\aa}C_{\!c\cc}C_{\!d\dd}=1}}\!\!\!
\frac{\SS_{\bb}\,S_d}{\SS_{\dd}\,S_b}\;\;
\I{a}{\aa}{\dd}{d}\;\I{b}{\bb}{\aa}{a}\;\I{b}{\bb'}{\cc}{c}\;
\I{c}{\cc}{\dd}{d}\;\W{a}{b}{c}{d}{u}\nonumber\eeqa
\setlength{\unitlength}{7mm}\begin{center}
\begin{picture}(3,5)
\thicklines\multiput(0.5,2.5)(1,1){2}{\line(1,-1){1}}
\multiput(0.5,2.5)(1,-1){2}{\line(1,1){1}}
\put(1.45,1.66){\spos{b}{\rightarrow}}
\put(1.5,2.5){\spos{}{u}}
\put(1.5,1.4){\spos{t}{\aa}}\put(2.6,2.5){\spos{l}{\bb}}
\put(1.5,3.6){\spos{b}{\cc}}\put(0.4,2.5){\spos{r}{\dd}}
\end{picture}
\text{1.5}{5}{0.9}{2.5}{}{\delta_{\bb\bb'}}
\text{1.8}{5}{0.9}{2.5}{}{=}
\text{2.5}{5}{0.9}{2.5}{}{\frac{\SS_{\bb}\,S_d}{\SS_{\dd}\,S_b}}
\begin{picture}(3,5)
\multiput(0.5,2.5)(1,1){2}{\line(1,-1){1}}
\multiput(0.5,2.5)(1,-1){2}{\line(1,1){1}}
\multiput(0.5,1.5)(0,2){2}{\line(1,0){2}}
\multiput(0.5,0.5)(0,0.15){7}{\line(0,1){0.1}}
\multiput(1.5,0.5)(0,0.15){7}{\line(0,1){0.1}}
\multiput(2.5,0.5)(0,0.15){7}{\line(0,1){0.1}}
\multiput(0.5,3.5)(0,0.15){7}{\line(0,1){0.1}}
\multiput(1.5,3.5)(0,0.15){7}{\line(0,1){0.1}}
\multiput(2.5,3.5)(0,0.15){7}{\line(0,1){0.1}}
\thicklines\multiput(0.5,0.5)(0,4){2}{\line(1,0){2}}
\multiput(0.5,1.74)(0,0.26){3}{\spos{}{.}}
\multiput(2.5,1.74)(0,0.26){3}{\spos{}{.}}
\multiput(0.5,2.74)(0,0.26){3}{\spos{}{.}}
\multiput(2.5,2.74)(0,0.26){3}{\spos{}{.}}
\multiput(-0.3,0.94)(0,0.26){13}{\spos{}{.}}
\put(0.1,0.6){\spos{}{.}}\put(-0.15,0.7){\spos{}{.}}
\put(0.1,4.4){\spos{}{.}}\put(-0.15,4.3){\spos{}{.}}
\multiput(1.5,1.5)(0,2){2}{\spos{}{\bullet}}
\multiput(0.5,1.5)(0,1){3}{\spos{}{\bullet}}
\multiput(1.44,3.6)(1,0){2}{\spos{br}{\nearrow}}
\multiput(1.44,1.49)(1,0){2}{\spos{tr}{\searrow}}
\put(1.45,1.66){\spos{b}{\rightarrow}}
\put(1.5,2.5){\spos{}{u}}
\put(0.45,0.45){\spos{tr}{\dd}}\put(1.5,0.4){\spos{t}{\aa}}
\put(2.55,0.45){\spos{tl}{\bb}}\multiput(2.6,1.5)(0,1){3}{\spos{l}{b}}
\put(2.55,4.55){\spos{bl}{\bb'}}\put(1.5,4.6){\spos{b}{\cc}}
\put(0.45,4.55){\spos{br}{\dd}}\multiput(0.35,1.5)(0,1){3}{\spos{r}{d}}
\put(1.94,1.57){\spos{b}{a}}\put(1.92,3.45){\spos{t}{c}}
\end{picture}\end{center}
for all $u$ and all $b$, $\aa$, $\bb$, $\bb'$, $\cc$, $\dd$ satisfying
$\AA_{\aa\bb}\:\AA_{\cc\bb'}\:\AA_{\dd\cc}\:\AA_{\dd\aa}
\:C_{\!b\bb}\:C_{\!b\bb'}=1$, and that the boundary weight intertwining
relation is equivalent to
\beq\eql{BWI}\BB{\aa}{\bb}{\cc}{u}\;=
\sum_{\stackrel{\scriptstyle ab}{\scriptscriptstyle 
A_{ba}A_{bc}C_{\!a\aa}C_{\!b\bb}=1}}\!\!\!
\frac{S_{b}\,\SS_{\cc}}{S_c\,\SS_{\bb}}\;\;
\I{a}{\aa}{\bb}{b}\;\I{c}{\cc}{\bb}{b}\;\B{a}{b}{c}{u}\eeq
\setlength{\unitlength}{7mm}\begin{center}
\begin{picture}(2,3)\thicklines\put(1.5,0.5){\line(0,1){2}}
\put(0.5,1.5){\line(1,-1){1}}\put(0.5,1.5){\line(1,1){1}}
\put(1,1.5){\spos{l}{u}}
\put(1.55,2.55){\spos{bl}{\aa}}\put(0.4,1.5){\spos{r}{\bb}}
\put(1.55,0.45){\spos{tl}{\cc}}
\end{picture}
\text{1.8}{3}{0.9}{1.5}{}{=}
\text{1.6}{3}{0.7}{1.5}{}{\frac{S_{b}\,\SS_{\cc}}{S_c\,\SS_{\bb}}}
\begin{picture}(3,3)
\put(1.5,0.5){\line(0,1){2}}\put(2.5,0.5){\line(0,1){2}}
\put(1.5,1.5){\line(1,-1){1}}\put(1.5,1.5){\line(1,1){1}}
\multiput(0.5,0.5)(0.15,0){7}{\line(1,0){0.1}}
\multiput(0.5,1.5)(0.15,0){7}{\line(1,0){0.1}}
\multiput(0.5,2.5)(0.15,0){7}{\line(1,0){0.1}}
\thicklines\put(0.5,0.5){\line(0,1){2}}
\multiput(1.74,0.5)(0.26,0){3}{\spos{}{.}}
\multiput(1.74,2.5)(0.26,0){3}{\spos{}{.}}
\multiput(1.5,1.5)(0,1){2}{\spos{}{\bullet}}\put(2.5,2.5){\spos{}{\bullet}}
\put(1.52,0.66){\spos{br}{\swarrow}}
\put(1.52,2.44){\spos{tr}{\nwarrow}}
\put(2,1.5){\spos{l}{u}}
\put(0.45,2.55){\spos{br}{\aa}}\put(0.4,1.5){\spos{r}{\bb}}
\put(0.45,0.45){\spos{tr}{\cc}}
\put(1.5,2.65){\spos{b}{a}}\put(2.6,2.6){\spos{bl}{a}}
\put(1.55,1.9){\spos{l}{b}}
\put(1.5,0.4){\spos{t}{c}}\put(2.55,0.45){\spos{tl}{c}}
\end{picture}\end{center}
for all $u$ and all $c$, $\aa$, $\bb$, $\cc$ satisfying
$\AA_{\bb\aa}\:\AA_{\bb\cc}\:C_{\!c\cc}=1$. We note that the
right sides of~\eqref{FWI1}, \eqref{FWI2} 
and~\eqref{BWI} must be independent of $d$, $b$ and $c$ respectively.

The importance of the intertwiner relations~\eqref{FIIR}--\eqref{BWIR}
is that they imply that if the Yang-Baxter equation and boundary Yang-Baxter
equation are satisfied by the weights of one model, then they are also
satisfied by the weights of the other model.  For if, on each side of the
Yang-Baxter equation for the weights of one
model, we introduce three intertwiner cells, apply~\eqref{FWIR} three 
times successively, introduce a further three intertwiner cells, and
apply~\eqref{FIIR} three times, then we obtain the Yang-Baxter equation
for the weights of the other model.  Similarly, if, on each side of the
boundary Yang-Baxter equation for the weights 
of one model, we introduce two intertwiner cells, apply~\eqref{FWIR} twice
and~\eqref{BWIR} twice, introduce a further two intertwiner cells, and 
apply~\eqref{SIIR} twice, then we obtain the boundary Yang-Baxter equation
for the weights of the other model.

\sect{$A_{\infty}$ Model}
\sub{Bulk Weights}
Throughout this and subsequent sections, we shall use the elliptic theta
functions,
\beq\begin{array}{rcl}
\rule[-3.5ex]{0ex}{3.5ex}\t(u,q)&=&\displaystyle2\,q^{\ssm{1/4}}\sin\!u\,
\prod_{n=1}^{\infty}\left(1-2\,q^{2n}\cos\!2u+q^{4n}\right)
\left(1-q^{2n}\right)\\
\rule[-3.5ex]{0ex}{3.5ex}\tt(u,q)&=&\displaystyle2\,q^{\ssm{1/4}}\cos\!u\,
\prod_{n=1}^{\infty}\left(1+2\,q^{2n}\cos\!2u+q^{4n}\right)
\left(1-q^{2n}\right)\\
\rule[-3.5ex]{0ex}{3.5ex}\ttt(u,q)&=&\displaystyle
\prod_{n=1}^{\infty}\left(1+2\,q^{2n-1}\cos\!2u+q^{4n-2}\right)
\left(1-q^{2n}\right)\\
\tttt(u,q)&=&\displaystyle\prod_{n=1}^{\infty}
\left(1-2\,q^{2n-1}\cos\!2u+q^{4n-2}\right)\left(1-q^{2n}\right)\;.
\end{array}\eeq
Since the nome $q$ is generally fixed, we shall abbreviate
\beq\th_i(u)=\th_i(u,q)\;.\eeq
We now consider the $A_{\infty}$ or eight-vertex solid-on-solid
model~\cite{Bax73b}.
The spins in this model take values from the set of all integers and 
the adjacency graph is
\setlength{\unitlength}{9mm}\beq
\raisebox{-0.5\unitlength}[0.5\unitlength][0.5\unitlength]{
\text{2.1}{1}{0}{0.5}{l}{A_{\infty}\;\;=}
\begin{picture}(7,1)
\put(1,0.5){\line(1,0){5}}
\multiput(0,0.5)(0.5,0){2}{\line(1,0){0.2}}
\multiput(6.2,0.5)(0.5,0){2}{\line(1,0){0.2}}
\multiput(1.5,0.5)(1,0){5}{\spos{}{\bullet}}
\put(1.4,0.35){\spos{t}{-2}}\put(2.4,0.35){\spos{t}{-1}}
\put(3.5,0.35){\spos{t}{0}}\put(4.5,0.35){\spos{t}{1}}
\put(5.5,0.35){\spos{t}{2}}\end{picture}
\text{0.3}{1}{0.3}{0.3}{b}{.}}\eeq
The bulk weights are
\beqa\rule[-3.5ex]{0ex}{3.5ex}
\W{a}{a\mipl1}{a}{a\plmi1}{u}&=&
\frac{\t(\l\mi u)}{\t(\l)}\noeqno
\rule[-3.5ex]{0ex}{3.5ex}\W{a\mipl1}{a}{a\plmi1}{a}{u}&=&
\left(\frac{\t((a\mi1)\l\pl\ww)\,\t((a\pl1)\l\pl\ww)}{\t(a\l)^2}
\right)^{\!\!\ssm{1\!/2}}\;\frac{\t(u)}{\t(\l)}\eql{AIFW}\\
\W{a\plmi1}{a}{a\plmi1}{a}{u}&=&\frac{\t(a\l\pl\ww\plmi u)}{\t(a\l\pl\ww)}
\nonumber\eeqa
where $\l$ and $\ww$ are arbitrary.

These weights satisfy the initial condition, reflection 
symmetry, crossing symmetry with crossing parameter $\l$ 
and crossing factors
\beq S_a=\t(a\l\pl\ww)\;,\eeq
the inversion relation with
\beq\eql{rho}\rho(u)=\frac{\t(\l\mi u)}{\t(\l)}\;,\eeq
and the Yang-Baxter equation.

The bulk weights for the critical $A_{\infty}$ model
are obtained by taking $q\rightarrow0$ so that all of
the $\t$ functions in~\eqref{AIFW} become $\sin$ functions.

\sub{Diagonal Boundary Weights}
Diagonal boundary weights for the $A_{\infty}$ model are
\beq\eql{AIDBW}\begin{array}{@{}l@{}}\rule[-3.5ex]{0ex}{3.5ex}
\makebox[146mm][l]{$\B{a}{a\plmi1}{a}{u}\;=\;
(x_1(a)\:\t(u)\:\t(u\mipl a\l\mipl\ww)\;+\;
x_2(a)\:\tttt(u)\:\tttt(u\mipl a\l\mipl\ww))\:f(a,u)$}\\
\B{a\plmi1}{a}{a\mipl1}{u}\;=\;0\end{array}\eeq 
where $x_1$, $x_2$ and $f$ are arbitrary.

We now prove that these boundary weights represent the general diagonal
solution of the boundary Yang-Baxter equation for the $A_{\infty}$ model.  
Having set $\smB{a\pm1}{a}{a\mipl1}{u}=0$, we 
find that the only spin assignments in the boundary Yang-Baxter 
equation~\eqref{BYBE} which
lead to non-trivial equations are $a=c=e$, $b=a\plmi1$, $d=a\mipl1$.  
These equations can be written
\beq\eql{AIBYBE}\BBB_a(u)^t\;\RRR_a(u,v)\;\BBB_a(v)\;=\;0\eeq
where
\[\BBB_a(u)\;=\;\left(\begin{array}{@{\,}c@{\,}}
\smB{a}{a+1}{a}{u}\\\smB{a}{a-1}{a}{u}\end{array}\right)\]
and
\[\RRR_a(u,v)\;=\;\left(\begin{array}{@{}r@{\quad}r@{\,}}
\rule[-1.5ex]{0ex}{1.5ex}
\t(u\mi v)\:\t(u\pl v\pl a\l\pl\ww)&-\t(u\pl v)\:\t(u\mi v\pl a\l\pl\ww)\\
-\t(u\pl v)\:\t(u\mi v\mi a\l\mi\ww)&\t(u\mi v)\:\t(u\pl v\mi a\l\mi\ww)
\end{array}\right)\;.\]
Decomposing each entry of $\RRR_a(u,v)$ using a standard elliptic addition
identity, we find that
\[\tttt(0)\;\tttt(a\l\pl\ww)\;\RRR_a(u,v)\;=\;\SSS_a(u)^t\;Q\;\SSS_a(v)\]
where
\[\SSS_a(u)\;=\;\left(\begin{array}{@{}r@{\quad}r@{\,}}
\rule[-1.5ex]{0ex}{1.5ex}
\tttt(u)\:\tttt(u\pl a\l\pl\ww)&-\tttt(u)\:\tttt(u\mi a\l\mi\ww)\\
-\t(u)\:\t(u\pl a\l\pl\ww)&\t(u)\:\t(u\mi a\l\mi\ww)\end{array}\right)
\quad\mbox{and}\quad
Q\;=\;\left(\begin{array}{@{}r@{\quad}c@{\,}}\rule[-1.5ex]{0ex}{1.5ex}0&1\\
-1&0\end{array}\right)\;.\]
We therefore obtain, from~\eqref{AIBYBE},
\beq\eql{Keq}K(u)^t\;Q\;K(v)\;=\;0\qquad\mbox{or}\qquad K_1(u)\:K_2(v)\,-\,
K_2(u)\:K_1(v)\;=\;0\eeq
\rule[-3ex]{0ex}{3ex}where $K(u)=\left(\begin{array}{@{}c@{}}
\rule[-1.5ex]{0ex}{1.5ex}K_1(u)\\K_2(u)\end{array}\right)=\SSS_a(u)\,
\BBB_a(u)$.
The general solution of~\eqref{Keq} is
$K(u)=\left(\begin{array}{@{}c@{}}\rule[-1.5ex]{0ex}{1.5ex}
x_1\\x_2\end{array}\right)f(u)$,
where $x_1$ and $x_2$ are arbitrary constants and $f$ is an arbitrary
function, and therefore the general solution of~\eqref{AIBYBE} is 
\begin{eqnarray*}\rule[-3.5ex]{0ex}{3.5ex}
\BBB_a(u)&=&\SSS_a(u)^{-1}\;\left(\begin{array}{@{\,}c@{\,}}
\rule[-1.5ex]{0ex}{1.5ex}x_1(a)\\x_2(a)\end{array}\right)\:\tilde{f}(a,u)\\
&=&\left(\begin{array}{@{}c@{\quad}c@{\,}}\rule[-1.5ex]{0ex}{1.5ex}
\t(u)\:\t(u\mi a\l\mi\ww)&\tttt(u)\:\tttt(u\mi a\l\mi\ww)\\
\t(u)\:\t(u\pl a\l\pl\ww)&\tttt(u)\:\tttt(u\pl a\l\pl\ww)\end{array}\right)
\:\left(\begin{array}{@{\,}c@{\,}}
\rule[-1.5ex]{0ex}{1.5ex}x_1(a)\\x_2(a)\end{array}\right)\:f(a,u)
\end{eqnarray*}
where $\tilde{f}(a,u)\:=\:\det\SSS _a(u)\;f(a,u)\:=\:-\tttt(0)\:\t(a\l\pl\ww)
\:\tttt(a\l\pl\ww)\:\t(2u)\:f(a,u)$, and $x_1$, $x_2$ and $f$ are arbitrary.
This solution matches~\eqref{AIDBW} and concludes our proof.

The $A_{\infty}$ diagonal boundary weights also satisfy
the boundary initial condition and the boundary inversion relation, with
$\chi$ and $\hat{\rho}$ given by~\eqref{DBIC} and~\eqref{DBIR}, 
and boundary crossing symmetry with
\beq\eql{AIBCS}\eta_a(u)=\frac{\t(2\l\mi2u)}{\t(\l)}\;
\frac{f(a,\l\mi u)}{f(a,u)}\;.\eeq
If we set
\beq\eql{Dxi}\begin{array}{@{}l@{}}\rule[-2.5ex]{0ex}{2.5ex}
\makebox[149mm][l]{$\bigl(x_1(a)\,,\,x_2(a)\bigr)=\rule[-2.5ex]{0ex}{2.5ex}
\textstyle\frac{1}{\tttt(0)\:\tttt(a\l\pl\ww)}\;
\Bigl(\tttt(\x_a)\:\tttt(a\l\pl\ww\pl\x_a)\,,\,
-\t(\x_a)\:\t(a\l\pl\ww\pl\x_a)\Bigr)$}\\
f(a,u)=1\end{array}\eeq
where $\x_a$ are arbitrary, then the weights
can be expressed in terms of $\t$ functions only as
\beq\B{a}{a\plmi1}{a}{u}=\t(u\plmi\x_a)
\:\t(u\mipl a\l\mipl\ww\mipl\x_a)\;.\eeq
The general diagonal solution of the boundary Yang-Baxter equation
for the critical $A_{\infty}$ model can be obtained 
from~\eqref{AIDBW} by
replacing $x_1(a)$ by $x_1(a)/q^{\ssm{1\!/2}}$ and 
taking $q\rightarrow0$, giving
\beq\B{a}{a\plmi1}{a}{u}=(x_1(a)\;\sin\!u\;\sin(u\mipl a\l\mipl\ww)+
x_2(a))\;f(a,u)\;.\eeq

\sub{Non-Diagonal Boundary Weights}
We now proceed to the case of non-diagonal boundary weights.  Using a method 
similar to that for the derivation of the diagonal boundary weights,
we have found that the general non-diagonal solution of the boundary 
Yang-Baxter equation for the $A_{\infty}$ model is
\beq\eql{AINDBW}\begin{array}{rcl}\rule[-4ex]{0ex}{4ex}
\B{a\plmi1}{a}{a\plmi1}{u}&=&K(a,\pm u)\;f^{\pa}(u)\\
\B{a\mipl1}{a}{a\plmi1}{u}&=&\kappa_{\pm}(a)\;k(u)\;f^{\pa}(u)\end{array}\eeq
where $\pa$ is the parity (even or odd) of $a$,
\beqa\lefteqn{\rule[-3ex]{0ex}{3ex}K(a,u)\;=\;
\frac{1}{\t(a\l\pl\ww)}\quad\times}\noeqno
&&\rule[-1.5ex]{0ex}{1.5ex}\Bigl(
x_1^{\pa}\;\t(u\pl a\l\pl\ww)\;\tt(u)\;\ttt(u)\;\tttt(u)\;+\;
x_2^{\pa}\;\tt(u\pl a\l\pl\ww)\;\t(u)\;\ttt(u)\;\tttt(u)\noeqno
&&\rule[-2ex]{0ex}{2ex}\qquad\mbox{}+\;
x_3^{\pa}\;\ttt(u\pl a\l\pl\ww)\;\t(u)\;\tt(u)\;\tttt(u)\;+\;
x_4^{\pa}\;\tttt(u\pl a\l\pl\ww)\;\t(u)\;\tt(u)\;\ttt(u)\Bigr)\noeqno
\eql{K}\lefteqn{\rule[-4ex]{0ex}{4ex}\qquad\quad\;\;=\;
\frac{\tt(0)\:\ttt(0)\:\tttt(0)\:\t(2u)}{2\:\t(a\l\pl\ww)}
\;\sum_{i=1}^{4}\;x_i^{\pa}\;
\frac{\th_i(u\pl a\l\pl\ww)}{\th_i(u)}}\\
\eql{kappa}\lefteqn{\rule[-2ex]{0ex}{2ex}\kappa_+(a)\;=\;\kappa_-(a)\;=\;
(K(a,-\sigma^{\pa})\;K(a,\sigma^{\pa}))^{\ssm{1\!/2}}}\\
\eql{k}\lefteqn{k(u)\;=\;\frac{\t(2u)}{\t(2\sigma^{\pa})}\;,}\eeqa
and $x_1$, $x_2$, $x_3$, $x_4$, $\sigma$ and $f$ are arbitrary.
In fact, $\kappa_-(a)$ and 
$\kappa_+(a)$ are only defined up to their product, which
must be $K(a,-\sigma^{\pa})\;K(a,\sigma^{\pa})$, however the 
choice~\eqref{kappa} has been made so that the weights satisfy reflection 
symmetry. These weights also satisfy the boundary initial condition with
\beq\eql{AINDBIC}\chi_{a+1}=
x_1^{\pa}\:\tt(0)\:\ttt(0)\:\tttt(0)\:f^{\pa}(0)\;,\eeq
boundary crossing symmetry with
\beq\eta_{a+1}(u)=\frac{\t(2\l\mi2u)}{\t(\l)}\;
\frac{f^{\pa}(\l\mi u)}{f^{\pa}(u)}\;,\eeq
and the boundary inversion relation with
\beqa\eql{AINDBIR}\lefteqn{\hat{\rho}_{a+1}(u)\;=}\\
&&\rule[-4.5ex]{0ex}{4.5ex}\t(\sigma^{\pa}\mi u)\;\t(\sigma^{\pa}\pl u)\,
\left(\left(\frac{\tt(u)\;\ttt(u)\;
\tttt(u)\;x_1^{\pa}}{\t(\sigma^{\pa})}\right)^{\!2}-
\left(\frac{\t(u)\;\ttt(u)\;
\tttt(u)\;x_2^{\pa}}{\tt(\sigma^{\pa})}\right)^{\!2}
\right.\noeqno&&\qquad\qquad\quad\left.
+\left(\frac{\t(u)\;\tt(u)\;
\tttt(u)\;x_3^{\pa}}{\ttt(\sigma^{\pa})}\right)^{\!2}
-\left(\frac{\t(u)\;\tt(u)\;
\ttt(u)\;x_4^{\pa}}{\tttt(\sigma^{\pa})}\right)^{\!2}
\right)\,f^{\pa}(-u)\;f^{\pa}(u)\,.\nonumber\eeqa
The $A_{\infty}$ non-diagonal boundary weights include those
which are obtained in~\cite{FanHouShi95} by using known boundary
weights for the eight-vertex model and vertex-face intertwiners.  

The weights can also be written in terms of 
\beq\tilde{\th}_i(u)=\th_i(u,q^2)\eeq
as
\beqa K(a,u)&=&\rule[-3.5ex]{0ex}{3.5ex}
y_1^{\pa}\;\frac{\tilde{\th}_1(2u\pl a\l\pl\ww)\:
\tilde{\th}_4(2u)}{\tilde{\th}_1(a\l\pl\ww)}\;+\;
y_2^{\pa}\;\frac{\tilde{\th}_4(2u\pl a\l\pl\ww)\:
\tilde{\th}_4(2u)}{\tilde{\th}_4(a\l\pl\ww)}\noeqno
&&\eql{K2}\rule[-3.5ex]{0ex}{3.5ex}\quad\mbox{}+\;
y_3^{\pa}\;\frac{\tilde{\th}_4(2u\pl a\l\pl\ww)\:
\tilde{\th}_1(2u)}{\tilde{\th}_1(a\l\pl\ww)}\;+\;
y_4^{\pa}\;\frac{\tilde{\th}_1(2u\pl a\l\pl\ww)\:
\tilde{\th}_1(2u)}{\tilde{\th}_4(a\l\pl\ww)}\\
k(u)&=&\frac{\tilde{\th}_1(2u)\:\tilde{\th}_4(2u)}
{\tilde{\th}_1(2\sigma^{\pa})\:\tilde{\th}_4(2\sigma^{\pa})}
\nonumber\eeqa
where 
\[\bigl(y_1^{\pa}\,,\,y_2^{\pa}\,,\,y_3^{\pa}\,,\,y_4^{\pa}\bigr)\;=\;
\textstyle\frac{\tt(0)\:\tilde{\th}_4(0)}{2}\;
\bigl(x_1^{\pa}+\,x_2^{\pa}\,,\,x_1^{\pa}-\,x_2^{\pa}\,,\,
x_4^{\pa}+\,x_3^{\pa}\,,\,x_4^{\pa}-\,x_3^{\pa}\bigr)\;,\]
or in terms of 
\beq\hat{\th}_i(u)=\th_i(u,q^{\ssm{1\!/2}})\eeq
as
\beqa K(a,u)&=&\rule[-3.5ex]{0ex}{3.5ex}
z_1^{\pa}\;\frac{\hat{\th}_1(u\pl\frac{a\l+\ww}{2})\:
\hat{\th}_2(u)}{\hat{\th}_1(\frac{a\l+\ww}{2})}\;+\;
z_2^{\pa}\;\frac{\hat{\th}_2(u\pl\frac{a\l+\ww}{2})\:
\hat{\th}_2(u)}{\hat{\th}_2(\frac{a\l+\ww}{2})}\noeqno
&&\rule[-3.5ex]{0ex}{3.5ex}\qquad\mbox{}+\;
z_3^{\pa}\;\frac{\hat{\th}_2(u\pl\frac{a\l+\ww}{2})\:
\hat{\th}_1(u)}{\hat{\th}_1(\frac{a\l+\ww}{2})}\;+\;
z_4^{\pa}\;\frac{\hat{\th}_1(u\pl\frac{a\l+\ww}{2})\:
\hat{\th}_1(u)}{\hat{\th}_2(\frac{a\l+\ww}{2})}\\
k(u)&=&\frac{\hat{\th}_1(u)\:\hat{\th}_2(u)}
{\hat{\th}_1(\sigma^{\pa})\:\hat{\th}_2(\sigma^{\pa})}
\nonumber\eeqa
where
\[\bigl(z_1^{\pa}\,,\,z_2^{\pa}\,,\,z_3^{\pa}\,,\,z_4^{\pa}\bigr)\;=\;
\textstyle\frac{\tttt(0)\:\hat{\th}_2(0)}{4}\;
\bigl(x_1^{\pa}+\,x_4^{\pa}\,,\,x_1^{\pa}-\,x_4^{\pa}\,,\,
x_3^{\pa}+\,x_2^{\pa}\,,\,x_3^{\pa}-\,x_2^{\pa}\bigr)\;.\]
Furthermore, if we set 
\beqa\eql{NDxi} x_i^{\pa}&=&
\frac{(-1)^{i+1}\:\th_i(\x_1^{\pa}\mi\n_1^{\pa})\:
\th_i(\x_1^{\pa}\pl\n_1^{\pa})\:\th_i(\x_2^{\pa}\pl\n_2^{\pa})\:
\th_i(\x_2^{\pa}\pl\n_2^{\pa})}{\tt(0)\:\ttt(0)\:\tttt(0)}\noeqno
\sigma^{\pa}&=&\x_1^{\pa}\pl\n_1^{\pa}\\
f^{\pa}(u)&=&1\nonumber\eeqa
where $\x_1^{\pa}$, $\n_1^{\pa}$, $\x_2^{\pa}$, $\n_2^{\pa}$ are arbitrary,
then the weights can be 
expressed in terms of $\t$ functions only as
\beqa\lefteqn{\rule[-3.5ex]{0ex}{3.5ex}\B{a\plmi1}{a}{a\plmi1}{u}\;=\;
\frac{1}{\t(a\l\pl\ww)\:\t(2\n_1^{\pa})}\quad\times}\noeqno
\lefteqn{\rule[-1.5ex]{0ex}{1.5ex}\textstyle\qquad\Bigl(
\t(\frac{a\l+\ww}{2}\pl\n_1^{\pa}\mi\n_2^{\pa})\:
\t(\frac{a\l+\ww}{2}\pl\n_1^{\pa}\pl\n_2^{\pa})\:
\t(\pm u\mi\x_1^{\pa}\pl\n_1^{\pa})\:
\t(\pm u\pl\x_1^{\pa}\pl\n_1^{\pa})\;\times}\noeqno
\lefteqn{\rule[-1.5ex]{0ex}{1.5ex}\textstyle\hspace{66mm}
\t(\pm u\pl\frac{a\l+\ww}{2}\mi\x_2^{\pa}\mi\n_1^{\pa})\:
\t(\pm u\pl\frac{a\l+\ww}{2}\pl\x_2^{\pa}\mi\n_1^{\pa})}\noeqno
\lefteqn{\rule[-1.5ex]{0ex}{1.5ex}\textstyle\qquad\quad\mbox{}-\;
\t(\frac{a\l+\ww}{2}\mi\n_1^{\pa}\mi\n_2^{\pa})\:
\t(\frac{a\l+\ww}{2}\mi\n_1^{\pa}\pl\n_2^{\pa})\:
\t(\pm u\mi\x_1^{\pa}\mi\n_1^{\pa})\:
\t(\pm u\pl\x_1^{\pa}\mi\n_1^{\pa})\;\times}\noeqno
\lefteqn{\rule[-2.5ex]{0ex}{2.5ex}\textstyle\hspace{72mm}
\t(\pm u\pl\frac{a\l+\ww}{2}\mi\x_2^{\pa}\pl\n_1^{\pa})\:
\t(\pm u\pl\frac{a\l+\ww}{2}\pl\x_2^{\pa}\pl\n_1^{\pa})\Bigr)}\noeqno
\\
\lefteqn{\rule[-3.5ex]{0ex}{3.5ex}\B{a\mipl1}{a}{a\plmi1}{u}\;=\;
\frac{\t(2u)}{\t(a\l\pl\ww)}\quad\times}\noeqno
\lefteqn{\rule[-1.5ex]{0ex}{1.5ex}\textstyle\quad\;\Bigl(-
\t(\frac{a\l+\ww}{2}\mi\n_1^{\pa}\mi\n_2^{\pa})\:
\t(\frac{a\l+\ww}{2}\mi\n_1^{\pa}\pl\n_2^{\pa})\:
\t(\frac{a\l+\ww}{2}\pl\n_1^{\pa}\mi\n_2^{\pa})\:
\t(\frac{a\l+\ww}{2}\pl\n_1^{\pa}\pl\n_2^{\pa})\;\times}\noeqno
&&\textstyle\quad\quad\!
\t(\frac{a\l+\ww}{2}\mi\x_1^{\pa}\mi\x_2^{\pa})\:
\t(\frac{a\l+\ww}{2}\mi\x_1^{\pa}\pl\x_2^{\pa})\:
\t(\frac{a\l+\ww}{2}\pl\x_1^{\pa}\mi\x_2^{\pa})\:
\t(\frac{a\l+\ww}{2}\pl\x_1^{\pa}\pl\x_2^{\pa})
\Bigr)^{\!\!\ssm{1\!/2}}\:.\nonumber\eeqa
The general non-diagonal solution of the boundary Yang-Baxter equation
for the critical $A_{\infty}$ model can be obtained
from~\eqref{K2} by 
replacing $y_4^{\pa}$ by $y_4^{\pa}/q$ and taking $q\rightarrow0$, giving
\beqa K(a,u)&\!=\!&\rule[-3.5ex]{0ex}{3.5ex}
y_1^{\pa}\:\frac{\sin(2u\pl a\l\pl\ww)}{\sin(a\l\pl\ww)}\;+\;
y_2^{\pa}\;+\;
y_3^{\pa}\:\frac{\sin(2u)}{\sin(a\l\pl\ww)}\;+\;
y_4^{\pa}\,\sin(2u\pl a\l\pl\ww)\,\sin(2u)\noeqno
k(u)&\!=\!&\frac{\sin(2u)}{\sin(2\sigma^{\pa})}\eeqa
which matches the trigonometric weights obtained 
in~\cite{AhnKoo96a}.

\sect{$A_L$ Models}
\sub{Bulk Weights}
We now consider the $A_L$, or Andrews-Baxter-Forrester, 
models~\cite{AndBaxFor84}, which can be regarded as restricted cases of the 
$A_{\infty}$ model.  There is one such model for each integer
$L\geq2$.  The spins in this model take values from the set 
$\{1,2,\ldots,L\}$ and the adjacency graph is
\setlength{\unitlength}{9mm}\beq\eql{AL}
\raisebox{-0.5\unitlength}[0.5\unitlength][0.5\unitlength]{
\text{2.1}{1}{0}{0.5}{l}{A_L\;\;=}\begin{picture}(6,1)
\multiput(0,0.5)(3.5,0){2}{\line(1,0){2.5}}
\multiput(2.7,0.5)(0.4,0){2}{\line(1,0){0.2}}
\multiput(0,0.5)(1,0){2}{\spos{}{\bullet}}
\multiput(5,0.5)(1,0){2}{\spos{}{\bullet}}
\put(0,0.35){\spos{t}{1}}\put(1,0.35){\spos{t}{2}}
\put(5,0.35){\spos{t}{L\mi1}}\put(6,0.35){\spos{t}{L}}\end{picture}
\text{0.3}{1}{0.3}{0.3}{b}{.}}\eeq
The $A_L$ bulk weights are obtained from~\eqref{AIFW} with $\ww=0$ as
\beqa\rule[-3.5ex]{0ex}{3.5ex}
\W{a}{a\mipl1}{a}{a\plmi1}{u}&=&
\frac{\t(\l\mi u)}{\t(\l)}\;,\qquad a=2,\ldots, L\mi1\noeqno
\rule[-3.5ex]{0ex}{3.5ex}\W{a\mipl1}{a}{a\plmi1}{a}{u}&=&
\left(\frac{\t((a\mi1)\l)\,\t((a\pl1)\l)}{\t(a\l)^2}
\right)^{\!\!\ssm{1\!/2}}\;\frac{\t(u)}{\t(\l)}\;,\qquad
a=2,\ldots, L\mi1\noeqno
\W{a\plmi1}{a}{a\plmi1}{a}{u}&=&\frac{\t(a\l\plmi u)}{\t(a\l)}
\;,\qquad\biggl\{\begin{array}{@{}l}+,\;a=1,\ldots, L\mi1\\
-,\;a=2,\ldots,L\end{array}\eql{ALFW}\eeqa
where
\beq\eql{lamA}\l\;=\;\frac{\pi}{L+\!1}\;.\eeq
These weights satisfy the Yang-Baxter equation, the initial condition, 
invariance under the symmetry transformation
\beq\eql{ALZS}Z(a)=L+1-a\;,\eeq
reflection symmetry, crossing symmetry with crossing parameter~\eqref{lamA} 
and crossing factors
\beq S_a=\t(a\l)\;,\eeq
and the inversion relation with $\rho$ given by~\eqref{rho}.

\sub{Diagonal Boundary Weights}
The general diagonal solution of the boundary Yang-Baxter equation
for the $A_L$ models is obtained from~\eqref{AIDBW} as
\beqa\eql{ALDBW}
\B{a}{a\plmi1}{a}{u}&=&(x_1(a)\;\t(u)\;\t(u\mipl a\l)+\noeqno
&&\raisebox{1ex}[1.5ex][0.5ex]{$
\qquad\qquad\qquad x_2(a)\;\tttt(u)\;\tttt(u\mipl a\l))\:f(a,u)
\;,\qquad\biggl\{\begin{array}{@{}l}+,\;a=1,\ldots, L\mi1\\
-,\;a=2,\ldots,L\end{array}$}\noeqno
\B{a\plmi1}{a}{a\mipl1}{u}&=&0\;,\qquad a=2,\ldots, L\mi1\eeqa
where $x_1$, $x_2$ and $f$ are arbitrary.

These weights satisfy the boundary initial condition, the boundary inversion
relation and boundary crossing symmetry, with
$\chi$, $\hat{\rho}$ and $\eta$ given 
by~\eqref{DBIC}, \eqref{DBIR} and~\eqref{AIBCS}.
The weights are also invariant under the transformation~\eqref{ALZS}
if the arbitrary parameters satisfy appropriate conditions, such as
\beq x_1(a)=-x_1(L\pl1\mi a)\;,\quad x_2(a)=x_2(L\pl1\mi a)\;,
\quad f(a,u)=f(L\pl1\mi a,u)\;.\eeq
By using~\eqref{Dxi} with $\ww=0$, we find that these weights match those
obtained in~\cite{BehPeaObr96}.

\sub{Non-Diagonal Boundary Weights}
To obtain non-diagonal boundary weights for the $A_L$ models, we set
\beq\eql{ALNDBW}\begin{array}{rcl}\rule[-4ex]{0ex}{4ex}
\B{a\plmi1}{a}{a\plmi1}{u}&=&K(a,\pm u)\;f^{\pa}(u)
\;,\qquad\biggl\{\begin{array}{@{}l}+,\;a=1,\ldots, L\mi1\\
-,\;a=2,\ldots,L\end{array}\\
\B{a\mipl1}{a}{a\plmi1}{u}&=&\kappa_{\pm}(a)\;k(u)\;f^{\pa}(u)
\;,\qquad a=2,\ldots, L\mi1\end{array}\eeq
where $f$ is arbitrary and $K$, $\kappa$ and $k$ are given 
by~\eqref{K}--\eqref{k} with $\ww=0$.
We then find that in order for the boundary Yang-Baxter equation
to be satisfied for spin assignments
which include the spin values 
$1$ or $L$, the constants $x_1$, $x_2$, $x_3$, $x_4$ and $\sigma$
must satisfy
\[\begin{array}{@{}c@{}}
\rule[-2ex]{0ex}{2ex}x_1^{\ev}+x_2^{\ev}+x_3^{\ev}+x_4^{\ev}=0\\
\rule[-2ex]{0ex}{2ex}x_1^{\pi_{\!L+1}}+x_2^{\pi_{\!L+1}}
-x_3^{\pi_{\!L+1}}-x_4^{\pi_{\!L+1}}=0\\
\begin{array}{@{}r@{}}\rule[-1.5ex]{0ex}{1.5ex}\Bigl(
\gamma^{\odd}_{1,-}x_1^{\odd}+\gamma^{\odd}_{2,-}x_2^{\odd}+
\gamma^{\odd}_{3,-}x_3^{\odd}+\gamma^{\odd}_{4,-}x_4^{\odd}\Bigr)\Bigl(
\gamma^{\odd}_{1,+}x_1^{\odd}+\gamma^{\odd}_{2,+}x_2^{\odd}+
\gamma^{\odd}_{3,+}x_3^{\odd}+\gamma^{\odd}_{4,+}x_4^{\odd}\Bigr)\quad\\=0
\end{array}\\
\Bigl(\gamma^{\pi_{\!L}}_{1,-}x_1^{\pi_{\!L}}+
\gamma^{\pi_{\!L}}_{2,-}x_2^{\pi_{\!L}}-
\gamma^{\pi_{\!L}}_{3,-}x_3^{\pi_{\!L}}-
\gamma^{\pi_{\!L}}_{4,-}x_4^{\pi_{\!L}}\Bigr)\Bigl(
\gamma^{\pi_{\!L}}_{1,+}x_1^{\pi_{\!L}}+
\gamma^{\pi_{\!L}}_{2,+}x_2^{\pi_{\!L}}-
\gamma^{\pi_{\!L}}_{3,+}x_3^{\pi_{\!L}}-
\gamma^{\pi_{\!L}}_{4,+}x_4^{\pi_{\!L}}
\Bigr)=0\end{array}\]
where
\[\gamma^{\pi}_{i,\pm}=
\frac{\th_i(\sigma^{\pi}\plmi\l)}{\th_i(\sigma^{\pi})}\;.\]
The $A_L$ non-diagonal boundary weights satisfy the boundary initial
condition, reflection symmetry, boundary crossing symmetry and the
boundary inversion relation, with
$\chi$, $\eta$ and $\hat{\rho}$ given 
by~\eqref{AINDBIC}--\eqref{AINDBIR}.
The weights are also invariant under the transformation~\eqref{ALZS}
if the arbitrary parameters satisfy appropriate conditions.

By using~\eqref{NDxi}, we find that these weights include those
obtained in~\cite{BehPea96}.

\sect{$D_L$ Models}
\sub{Bulk Weights}
We now consider the $D_L$ models~\cite{Pas87b}. There is one such model for 
each integer $L\geq3$.  The spins in this model take values from the set 
$\{1,2,\ldots,L\}$ and the adjacency graph is
\setlength{\unitlength}{9mm}\beq\eql{DL}
\raisebox{-1\unitlength}[1\unitlength][1\unitlength]{
\text{2.1}{2}{0}{1}{l}{D_L\;\;=}
\begin{picture}(7.5,2)
\multiput(0,1)(3.5,0){2}{\line(1,0){2.5}}
\multiput(2.7,1)(0.4,0){2}{\line(1,0){0.2}}
\put(6,1){\line(2,1){1}}\put(6,1){\line(2,-1){1}}
\multiput(0,1)(1,0){2}{\spos{}{\bullet}}
\multiput(5,1)(1,0){2}{\spos{}{\bullet}}
\multiput(7,0.5)(0,1){2}{\spos{}{\bullet}}
\put(0,0.85){\spos{t}{1}}\put(1,0.85){\spos{t}{2}}
\put(4.95,0.85){\spos{t}{L\mi3}}\put(5.85,0.85){\spos{t}{L\mi2}}
\put(7,1.63){\spos{b}{L\mi1}}\put(7,0.37){\spos{t}{L}}
\end{picture}\text{0.3}{2}{0.3}{0.8}{b}{.}}\eeq
The $D_L$ bulk weights are 
\beq\eql{DLFW}\begin{array}{@{\protect\rule[-4ex]{0ex}{4ex}}l}
\W{a}{a\mipl1}{a}{a\plmi1}{u}=\displaystyle
\frac{\t(\l\mi u)}{\t(\l)}\;,\qquad a=2,\ldots,L\mi3\\
\W{L\mi2}{L\mi3}{L\mi2}{L\mi1}{u}=\W{L\mi2}{L\mi1}{L\mi2}{L\mi3}{u}\\
\qquad\qquad=\W{L\mi2}{L\mi3}{L\mi2}{L}{u}=\W{L\mi2}{L}{L\mi2}{L\mi3}{u}=
\displaystyle\frac{\t(\l\mi u)}{\t(\l)}\\
\W{L\mi2}{L\mi1}{L\mi2}{L}{u}=\W{L\mi2}{L}{L\mi2}{L\mi1}{u}=
\displaystyle\frac{\t(\l)\:\tt(u)-\tt(\l)\:\t(u)}{\tt(0)\:\t(\l)}\\
\W{a\mipl1}{a}{a\plmi1}{a}{u}=\displaystyle\left(\frac{\t((a\mi1)\l)\,
\t((a\pl1)\l)}{\t(a\l)^2}\right)^{\!\!\ssm{1\!/2}}\;\frac{\t(u)}{\t(\l)}
\;,\qquad a=2,\ldots, L\mi3\\
\W{L\mi3}{L\mi2}{L\mi1}{L\mi2}{u}=\W{L\mi1}{L\mi2}{L\mi3}{L\mi2}{u}\\
\qquad\qquad=\W{L\mi3}{L\mi2}{L}{L\mi2}{u}=\W{L}{L\mi2}{L\mi3}{L\mi2}{u}=
\displaystyle\left(\frac{\tt(0)\,\tt(2\l)}{2\,\tt(\l)^2}\right)^{\!\!
\ssm{1\!/2}}\;\frac{\t(u)}{\t(\l)}\\
\W{L\mi1}{L\mi2}{L}{L\mi2}{u}=\W{L}{L\mi2}{L\mi1}{L\mi2}{u}=\displaystyle
\frac{1}{2}\left(
\frac{\tt(\l\mi u)}{\tt(\l)}-\frac{\t(\l\mi u)}{\t(\l)}\right)\\
\W{a\plmi1}{a}{a\plmi1}{a}{u}=\displaystyle\frac{\t(a\l\plmi u)}{\t(a\l)}
\;,\qquad\biggl\{\begin{array}{@{}l}+,\;a=1,\ldots, L\mi3\\
-,\;a=2,\ldots,L\mi2\end{array}\\
\W{L\mi1}{L\mi2}{L\mi1}{L\mi2}{u}=\W{L}{L\mi2}{L}{L\mi2}{u}=\displaystyle
\frac{1}{2}\left(
\frac{\tt(\l\mi u)}{\tt(\l)}+\frac{\t(\l\mi u)}{\t(\l)}\right)\\
\W{L\mi2}{L\mi1}{L\mi2}{L\mi1}{u}=\W{L\mi2}{L}{L\mi2}{L}{u}=
\displaystyle\frac{\t(\l)\:\tt(u)+\tt(\l)\:\t(u)}{\tt(0)\:\t(\l)}
\end{array}\eeq
where
\beq\eql{lamD}\l\;=\;\frac{\pi}{2(L-1)}\;.\eeq
These weights satisfy the Yang-Baxter equation, the initial condition, 
invariance under the symmetry transformation
\beq\eql{DLZS}Z(a)=\left\{\begin{array}{@{\,}c@{\,,\quad}l}a&
a=1,\ldots,L\mi2\\
L&a=L\mi1\\L\mi1&a=L\;,\end{array}\right.\eeq
reflection symmetry, crossing symmetry with crossing parameter~\eqref{lamD}
and crossing factors
\beq S_a=\left\{\begin{array}{@{\,}l@{\,,\quad}l}\rule[-1ex]{0ex}{1ex}
\t(a\l)&a=1,\ldots,L\mi2\\
\tt(0)/2&a=L\mi1,L\;,\end{array}\right.\eeq
and the inversion relation with $\rho$ given by~\eqref{rho}.

\sub{Diagonal Boundary Weights}
We have found, using a method similar to
that for the $A_L$ models, that the general 
diagonal solution of the boundary
Yang-Baxter equation for the $D_L$ models is
\beq\eql{DLDBW}\begin{array}{@{}l}
\B{a}{a\plmi1}{a}{u}=(x_1(a)\;\t(u)\;\t(u\mipl a\l)+\\
\raisebox{0.5ex}[1.5ex][3.5ex]{$\qquad\qquad\qquad\qquad\qquad\quad
x_2(a)\;\tttt(u)\;\tttt(u\mipl a\l))\;f(a,u)
\;,\qquad\biggl\{\begin{array}{@{}l}+,\;a=1,\ldots, L\mi3\\
-,\;a=2,\ldots,L\mi3\end{array}$}\\
\rule[-4ex]{0ex}{4ex}\B{L\mi2}{a}{L\mi2}{u}=k_i(a,u)\;f(L\mi2,u)\;,\qquad
a=L\mi3,L\mi1,L\\
\rule[-4ex]{0ex}{4ex}\B{a}{L\mi2}{a}{u}=
\ttt(u)\;\tttt(u)\;f(a,u)\;,\qquad a=L\mi1,L\\
\rule[-4ex]{0ex}{4ex}\B{a\plmi1}{a}{a\mipl1}{u}=0\;,\qquad
a=2,\ldots,L\mi2\\
\B{L\mi3}{L\mi2}{L}{u}\!=\!\B{L}{L\mi2}{L\mi3}{u}
\!=\!\B{L\mi1}{L\mi2}{L}{u}\!=\!\B{L}{L\mi2}{L\mi1}{u}\!\!=\!0
\end{array}\eeq
where
\beq\begin{array}{@{}l}\rule[-2ex]{0ex}{2ex}
k_1(L\mi3,u)\;=\;x_1(L\mi2)\;\t(u)\;\tt(u\mi\l)+
x_2(L\mi2)\;\tttt(u)\;\ttt(u\mi\l)\\
\rule[-2ex]{0ex}{2ex}k_1(L\mi1,u)\;=\;k_1(L,u)\;=\;k_1(L\mi3,-u)\\
\rule[-2ex]{0ex}{2ex}k_2(L\mi3,u)\;=\;\Bigl(x_1(L\mi2)\;\ttt(u\mi\l)-
x_2(L\mi2)\;\tttt(u\mi\l)\Bigr)\;\times\\
\rule[-2ex]{0ex}{2ex}\hspace{47mm}\Bigl(x_1(L\mi2)\;\ttt(u\mi\l)+
x_2(L\mi2)\;\tttt(u\mi\l)\Bigr)\;\ttt(u)\;\tttt(u)\\
\rule[-2ex]{0ex}{2ex}k_2(L\mi1,u)\;=\;\Bigl(x_1(L\mi2)\;\ttt(u\mi\l)-
x_2(L\mi2)\;\tttt(u\mi\l)\Bigr)\;\times\\
\rule[-2ex]{0ex}{2ex}\hspace{47mm}\Bigl(x_1(L\mi2)\;\ttt(u\pl\l)+
x_2(L\mi2)\;\tttt(u\pl\l)\Bigr)\;\ttt(u)\;\tttt(u)\\
k_2(L,u)\;=\;k_2(L\mi1,-u)\;,\end{array}\eeq
$x_1$, $x_2$ and $f$ are arbitrary,
and $i$ may be chosen arbitrarily as $1$ or $2$.

These weights satisfy the boundary initial condition, the boundary inversion
relation and boundary crossing symmetry, with
$\chi$, $\hat{\rho}$ and $\eta$  given 
by~\eqref{DBIC}, \eqref{DBIR} and~\eqref{AIBCS}.
The weights are also invariant under the transformation~\eqref{DLZS} 
if the arbitrary parameters satisfy appropriate conditions, such as 
\beq i=1\;,\qquad f(L\mi1,u)=f(L,u)\;.\eeq

\sub{$A_{2L-3}$--$D_L$ Intertwiner}
The $A_{2L-3}$ and $D_L$ models can be related by intertwiner
cells~\cite{FenGin89,DifZub90,Roc90,Dif92,PeaZho93}. 
The intertwiner graph, whose adjacency matrix satisfies 
the intertwining relation, is 
\setlength{\unitlength}{9mm}\begin{center}
\begin{picture}(10,4)
\multiput(1,0.5)(0,3){2}{\line(1,0){2.5}}
\multiput(3.7,0.5)(0,3){2}{\line(1,0){0.2}}
\multiput(4.1,0.5)(0,3){2}{\line(1,0){0.2}}
\multiput(4.5,0.5)(0,3){2}{\line(1,0){3.5}}\put(8,2){\oval(3,3)[r]}
\multiput(1,0.5)(1,0){2}{\line(0,1){0.4}}
\multiput(1,1.05)(1,0){2}{\vector(0,1){0.4}}
\multiput(1,1.6)(1,0){2}{\line(0,1){0.8}}
\multiput(1,2.95)(1,0){2}{\vector(0,-1){0.4}}
\multiput(1,3.5)(1,0){2}{\line(0,-1){0.4}}
\multiput(6,0.5)(1,0){2}{\line(0,1){0.4}}
\multiput(6,1.05)(1,0){2}{\vector(0,1){0.4}}
\multiput(6,1.6)(1,0){2}{\line(0,1){0.8}}
\multiput(6,2.95)(1,0){2}{\vector(0,-1){0.4}}
\multiput(6,3.5)(1,0){2}{\line(0,-1){0.4}}
\multiput(9.5,2)(-1.1,0.3667){2}{\line(-3,1){0.4}}
\put(8.95,2.1833){\vector(-3,1){0.4}}
\multiput(9.5,2)(-1.1,-0.3667){2}{\line(-3,-1){0.4}}
\put(8.95,1.8167){\vector(-3,-1){0.4}}
\thicklines
\multiput(1,2)(3.5,0){2}{\line(1,0){2.5}}
\multiput(3.7,2)(0.4,0){2}{\line(1,0){0.2}}
\put(7,2){\line(2,1){1}}\put(7,2){\line(2,-1){1}}
\multiput(1,0.5)(0,1.5){3}{\spos{}{\bullet}}
\multiput(2,0.5)(0,1.5){3}{\spos{}{\bullet}}
\multiput(6,0.5)(0,1.5){3}{\spos{}{\bullet}}
\multiput(7,0.5)(0,1.5){3}{\spos{}{\bullet}}
\put(9.5,2){\spos{}{\bullet}}
\multiput(8,1.5)(0,1){2}{\spos{}{\bullet}}
\put(1,3.65){\spos{b}{1}}\put(2,3.65){\spos{b}{2}}
\put(6,3.65){\spos{b}{L\mi3}}\put(7,3.65){\spos{b}{L\mi2}}
\put(9.6,2){\spos{l}{L\mi1}}
\put(7,0.35){\spos{t}{L}}\put(6,0.35){\spos{t}{L\pl1}}
\put(2,0.35){\spos{t}{2L\mi4}}\put(0.9,0.35){\spos{t}{2L\mi3}}
\put(1.07,1.92){\spos{tl}{1}}\put(2.07,1.92){\spos{tl}{2}}
\put(5.93,1.92){\spos{tr}{L\mi3}}\put(6.93,1.92){\spos{tr}{L\mi2}}
\put(8,2.63){\spos{b}{L\mi1}}\put(8,1.37){\spos{t}{L}}
\put(0.75,3.5){\pos{r}{A_{2L\mi3}}}\put(0.75,2){\pos{r}{D_L}}
\end{picture}\end{center}
and the intertwiner cells are
\beq\eql{DLIC}
\I{a}{a}{a\plmi1}{a\plmi1}=\I{2L\mi2\mi a}{a}{a\plmi1}{2L\mi2\mipl1}=1
\;,\qquad\biggl\{\begin{array}{@{}l}+,\;a=1,\ldots, L\mi3\\
-,\;a=2,\ldots,L\mi2\end{array}\quad\qquad\qquad\qquad\eeq
\[\begin{array}{@{}l}\rule[-5ex]{0ex}{5ex}\displaystyle
\I{L}{L\mi2}{L\mi1}{L\mi1}=\I{L}{L\mi2}{L}{L\mi1}=
\I{L\mi2}{L\mi2}{L\mi1}{L\mi1}=\frac{1}{\!\!\sqrt{2}}\;,\quad
\I{L\mi2}{L\mi2}{L}{L\mi1}=-\frac{1}{\!\!\sqrt{2}}\\
\I{L\mi1}{L\mi1}{L\mi2}{L}=\I{L\mi1}{L}{L\mi2}{L}=
\I{L\mi1}{L\mi1}{L\mi2}{L\mi2}=1\:,\quad
\I{L\mi1}{L}{L\mi2}{L\mi2}=-1\;.
\end{array}\]
These intertwiner cells satisfy the intertwiner inversion 
relations and the bulk weight 
intertwining relation. They also satisfy 
the boundary weight intertwining relation together with 
the $A_{2L-3}$ and $D_L$ diagonal boundary weights, provided that the
arbitrary parameters satisfy appropriate conditions, such as
\beq\begin{array}{l}\rule[-2.5ex]{0ex}{2.5ex}
i=1\,,\quad x^A_1(a)=-x^A_1(2L\mi2\mi a)=x^D_1(a)\,,\quad 
x^A_2(a)=x^A_2(2L\mi2\mi a)=x^D_2(a)\,,\\
\rule[-2.5ex]{0ex}{2.5ex}\qquad\qquad\qquad\qquad
f^A(a,u)=f^A(2L\mi2\mi a,u)=f^D(a,u)\,,\qquad a=1,\ldots,L\mi2\\
\quad x^A_1(L\mi1)=0\,,\quad
x^A_2(L\mi1)\,f^A(L\mi1,u)=f^D(L\mi1,u)=f^D(L,u)\;.\end{array}\eeq

\sect{Temperley-Lieb Models}
\sub{Bulk Weights}
We now consider the Temperley-Lieb models~\cite{OwcBax87,Pas87a}.
The adjacency graph $\G$ for these models can be any finite, connected 
graph which has only bidirectional, single bonds.  By the Perron-Frobenius
theorem, the adjacency matrix of $\G$ has a unique, positive maximum
eigenvalue $\Lambda$, with an associated eigenvector $(S_1,S_2,\ldots)$ 
which has all positive entries.

The Temperley-Lieb bulk weights are
\beq\eql{TLFW}
\W{a}{b}{c}{d}{u}\;=\;\frac{s(\l\mi u)}{s(\l)}\;\delta_{ac}\;+\;
\frac{\;(S_a\,S_c)^{\ssm{1\!/2}}}{S_b}\;
\frac{s(u)}{s(\l)}\;\delta_{bd}\eeq
where $\l$ is any solution of
\beq\eql{lamTL}\Lambda\;=\;2\,c(\l)\eeq
and
\beq\eql{TLsc}s(u)\;=\;\left\{\begin{array}{@{\,}l@{\quad}l}
\sin\!u\,,&\Lambda<2\\
u\,,&\Lambda=2\\
\sinh\!u\,,&\Lambda>2\end{array}\right.\;,\qquad\qquad
c(u)\;=\;\left\{\begin{array}{@{\,}l@{\quad}l}
\cos\!u\,,&\Lambda<2\\
1\,,&\Lambda=2\\
\cosh\!u\,,&\Lambda>2\;.\end{array}\right.\eeq
These weights satisfy the Yang-Baxter equation, the initial condition, 
reflection symmetry, crossing symmetry with crossing 
parameter~$\l$ and crossing factors $S_a$,
and the inversion relation with 
\beq\rho(u)=\frac{s(\l\mi u)}{s(\l)}\;.\eeq
{}From~\eqref{FTM} and~\eqref{TLFW}, we obtain the Temperley-Lieb bulk
face transfer matrices,
\beq\eql{TLFTM}X_j(u)\;=\;\frac{s(\l\mi u)}{s(\l)}\;I\;+\;
\frac{s(u)}{s(\l)}\;e_j\eeq
where 
\beq e_{j\:\raisebox{-0.2ex}{$\scriptstyle(a_0\ldots a_{\N\pl1}),
(b_0\ldots b_{\N\pl1})$}}\;\;=\;\;\prod_{k=0}^{j\mi1}\delta_{a_kb_k}\;
\frac{\;(S_{a_j}\,S_{b_j})^{\ssm{1\!/2}}}{S_{a_{j-1}}}\;
\delta_{a_{j-1}a_{j+1}}\;\prod_{k=j\pl1}^{\N\pl1}\delta_{a_kb_k}\;.\eeq
The matrices $e_j$ satisfy
\beqa e_i\:e_j\,-\,e_j\:e_i&=&0\,,\quad|i\mi j|>1\noeqno
\eql{TLA}e_j\:e_{j\pm1}\:e_j&=&e_j\\
e_j^2&=&\Lambda\:e_j\nonumber\eeqa
and therefore form a representation of the Temperley-Lieb 
algebra~\cite{TemLie71}.
In fact, it can be shown straightforwardly that if $e_j$ are any 
matrices which satisfy~\eqref{TLA},
then $X_j(u)$, defined by~\eqref{lamTL},~\eqref{TLsc} 
and~\eqref{TLFTM} alone, satisfy the Yang-Baxter equation~\eqref{MYBE}. We
also note that~\eqref{TLFTM} immediately implies commutation of the bulk
face transfer matrices,
\beq\eql{FTMcomm}X_j(u)\:X_j(v)=X_j(v)\:X_j(u)\;.\eeq
It is known that the only simple, connected graphs with $\Lambda<2$
are $A_L$, $D_L$, and $E_6$, $E_7$ and $E_8$,
\setlength{\unitlength}{9mm}\beq\eql{EL}
\begin{array}{c}\text{2}{2}{0}{0.5}{l}{E_6\;\;=}
\begin{picture}(6.5,2)
\put(0,0.5){\line(1,0){4}}\put(2,0.5){\line(0,1){1}}
\multiput(0,0.5)(1,0){5}{\spos{}{\bullet}}\put(2,1.5){\spos{}{\bullet}}
\put(0,0.35){\spos{t}{1}}\put(1,0.35){\spos{t}{2}}
\put(2,0.35){\spos{t}{3}}\put(3,0.35){\spos{t}{4}}
\put(4,0.35){\spos{t}{5}}\put(2,1.65){\spos{b}{6}}
\put(4.6,0.4){\pos{t}{,}}
\end{picture}
\text{2}{2}{0}{0.5}{l}{E_7\;\;=}
\begin{picture}(5,2)
\put(0,0.5){\line(1,0){5}}\put(3,0.5){\line(0,1){1}}
\multiput(0,0.5)(1,0){6}{\spos{}{\bullet}}\put(3,1.5){\spos{}{\bullet}}
\put(0,0.35){\spos{t}{1}}\put(1,0.35){\spos{t}{2}}
\put(2,0.35){\spos{t}{3}}\put(3,0.35){\spos{t}{4}}
\put(4,0.35){\spos{t}{5}}\put(5,0.35){\spos{t}{6}}
\put(3,1.65){\spos{b}{7}}
\end{picture}\\
\text{2}{2}{0}{0.5}{l}{E_8\;\;=}
\begin{picture}(9,2.3)
\put(0,0.5){\line(1,0){6}}\put(4,0.5){\line(0,1){1}}
\multiput(0,0.5)(1,0){7}{\spos{}{\bullet}}\put(4,1.5){\spos{}{\bullet}}
\put(0,0.35){\spos{t}{1}}\put(1,0.35){\spos{t}{2}}
\put(2,0.35){\spos{t}{3}}\put(3,0.35){\spos{t}{4}}
\put(4,0.35){\spos{t}{5}}\put(5,0.35){\spos{t}{6}}
\put(6,0.35){\spos{t}{7}}\put(4,1.65){\spos{b}{8}}
\put(6.5,0.3){\pos{b}{.}}
\end{picture}\end{array}\eeq
It can be shown that for $A_L$ and $D_L$, $\l$ is given 
by~\eqref{lamA} and~\eqref{lamD}, while for $E_{6,7,8}$,
\beq\l\;=\;\left\{\begin{array}{@{\,}l@{\,,\quad}l}
\pi/12&E_6\\
\pi/18&E_7\\
\pi/30&E_8\;.\end{array}\right.\eeq
For these graphs, it is also known that we may set
\beq\eql{TLS}S_a\;\;=\;\;\left\{\begin{array}{@{}ll}
\rule[-2.5ex]{0ex}{2.5ex}\;\;\:\sin\!a\l\,,\;\;\,a=1,\ldots,L\;;&A_L\\
\rule[-4.5ex]{0ex}{4.5ex}\left\{\begin{array}{@{}ll}
\rule[-1.5ex]{0ex}{1.5ex}\sin\!a\l\,,&a=1,\ldots,L\mi2\\
1/2\,,&a=L\mi1,L\end{array}\right.\;;&D_L\\
\left\{\begin{array}{@{}ll}
\rule[-1.5ex]{0ex}{1.5ex}\sin\!a\l\,,&a=1,\ldots,L\mi3\\
\rule[-1.5ex]{0ex}{1.5ex}\sin(L\mi1)\l\,/\,(2\cos\!\l)\,,&a=L\mi2\\
\rule[-1.5ex]{0ex}{1.5ex}2\cos(L\mi2)\l\,\sin\!\l\,,&a=L\mi1\\
\sin(L\mi3)\l\,/\,(2\cos\!\l)\,,&a=L\end{array}\right.\;;&E_L\,,\;L=6,7,8\;.
\end{array}\right.\eeq
{}From~\eqref{TLFW} and~\eqref{TLS} we find that
the bulk weights for the Temperley-Lieb models 
with adjacency graphs $A_L$ and $D_L$ match those for the 
critical $A_L$ and $D_L$ models, obtained by
taking $q\rightarrow0$ in~\eqref{ALFW} and~\eqref{DLFW}.

\sub{Diagonal Boundary Weights}
Diagonal boundary weights for the Temperley-Lieb models are
\beqa\eql{TLDBW}\lefteqn{\B{a}{b}{c}{u}\;=}
\rule[-3.5ex]{0ex}{3.5ex}\qquad\\
&&\left\{\begin{array}{@{\,}r@{\,,\quad}l}
\rule[-4.5ex]{0ex}{4.5ex}\displaystyle
\Bigl[x_1(a)\,s(u)\,\Bigl(s(u\pl\l)\,-\!\!\sum_{d\,\in\,\nu(a)}\!\!
S_d/S_a\,s(u)\Bigr)\:+\:x_2(a)\Bigr]
\:f(a,u)\;\delta_{ac}&b\in\nu(a)\\
\displaystyle\Bigl[-x_1(a)\,s(u)\,\Bigl(s(u\pl\l)\,
-\!\!\sum_{d\,\in\,\nu'(a)}\!\!S_d/S_a\,s(u)\Bigr)\:+\:x_2(a)\Bigr]
\:f(a,u)\;\delta_{ac}&b\in\nu'(a)
\end{array}\right.\nonumber\eeqa
where, for each $a$, $\nu(a)$ and $\nu'(a)$ are any non-intersecting
sets whose union is the set of neighbours of $a$,
and $x_1$, $x_2$ and $f$ are arbitrary.

We now prove that these boundary weights represent the general
diagonal solution of the boundary Yang-Baxter equation for the 
Temperley-Lieb models.
Having set $\smB{a}{b}{c}{u}=\BBB_a(b,u)\,\delta_{ac}$,
we find, using the Temperley-Lieb bulk weights, that
the only spin assignments in the boundary Yang-Baxter 
equation~\eqref{BYBE} which lead to non-trivial equations are
those in which $a=c=e$, and $b$ and $d$ are distinct
neighbours of $a$.
These equations are
\[\begin{array}{@{}r@{}}\rule[-2ex]{0ex}{2ex}
0\;=\;\E_a(b,d)\;=\;s(u\pl v)\,s(u\mi v\mi\l)\,
(\BBB_a(b,u)\,\BBB_a(d,v)-\BBB_a(d,u)\,\BBB_a(b,v))\\
\rule[-1.5ex]{0ex}{1.5ex}-\;s(u\mi v)\,s(u\pl v\mi\l)\,
(\BBB_a(b,u)\,\BBB_a(b,v)-\BBB_a(d,u)\,\BBB_a(d,v))\\
\displaystyle+\;s(u\mi v)\,s(u\pl v)\!\sum_{c\,\in\,\NN(a)}\!
S_c\,\BBB_a(c,u)/S_a\,(\BBB_a(b,v)-\BBB_a(d,v))\end{array}\]
where $\NN(a)$ is the set of neighbours of $a$.

We shall from now on treat $a$ as fixed.
If $a$ has $n$ neighbours, then
there are, since $\E_a(b,d)=-\,\E_a(d,b)$,
$n(n\mi 1)/2$ distinct equations for the $n$ boundary weights
$\BBB_a(b,u)$.

Throughout this proof, we shall also use the eigenvector equation
\[\sum_{b\,\in\,\NN(a)}S_b/S_a\;=\;2\,c(\l)\;.\]
We now observe that
\begin{eqnarray*}\lefteqn{0\;=\;\E_a(b,c)\,+\,\E_a(c,d)\,+\,
\E_a(d,b)\;=}\qquad\qquad\qquad\\
&&s(u\pl v)\,s(u\mi v\mi\l)\,\det
\left(\begin{array}{ccc}1&1&1\\
\rule[-1.5ex]{0ex}{1.5ex}\BBB_a(b,u)&\BBB_a(c,u)&\BBB_a(d,u)\\
\BBB_a(b,v)&\BBB_a(c,v)&\BBB_a(d,v)\end{array}\right)\;.\end{eqnarray*}
The general solution of this system of equations is
\[\BBB_a(b,u)\;=\;y(b)\,g(u)\;+\;h(u)\]
where $y(b)$ are arbitrary constants and $g$ and $h$ are 
arbitrary functions.
Using this solution, we obtain
\begin{eqnarray*}\rule[-2.5ex]{0ex}{2.5ex}
\lefteqn{0\;=\;\E_a(b,d)\;=\;(y(d)-y(b))\,\Bigl(
s(\l)\,s(2v)\,g(u)\,h(v)\,-\,s(\l)\,s(2u)\,h(u)\,g(v)}
\qquad\\
&&\mbox{}+\,s(u\mi v)\,\Bigl(
(y(b)+y(d))\,s(u\pl v\mi\l)\,-\!
\sum_{c\,\in\,\NN(a)}y(c)\,S_c/S_a\,s(u\pl v)
\Bigr)\,g(u)\,g(v)\Bigr)\;.\end{eqnarray*}
These equations are satisfied if $y(b)$ are equal for all
$b\in\NN(a)$.\rule[-1.5ex]{0ex}{1.5ex}
In order to obtain the remaining solutions, we assume 
$y(\tilde{b})\neq y(\tilde{d})$ for particular 
$\tilde{b}$ and $\tilde{d}$.
We now transform
\begin{eqnarray*}\rule[-2ex]{0ex}{2ex}
g(u)&=&s(\l)\,s(2u)\,\tilde{g}(u)\\
h(u)&=&\Bigl(
(y(\tilde{b})+y(\tilde{d}))\,s(u)\,s(u\mi\l)\,-\!
\sum_{c\,\in\,\NN(a)}y(c)\,S_c/S_a\,s(u)^2
\Bigr)\,\tilde{g}(u)\;+\;\tilde{h}(u)\end{eqnarray*}
with $\tilde{g}$ and $\tilde{h}$ arbitrary, which gives
\begin{eqnarray*}\rule[-2ex]{0ex}{2ex}
\lefteqn{0\;=\;\E_a(b,d)\;=\;\bigl(y(d)-y(b)\bigr)\,
s(\l)^2\,s(2u)\,s(2v)\,\bigl(\tilde{g}(u)\,\tilde{h}(v)\;-
\;\tilde{h}(u)\,\tilde{g}(v)}\qquad\qquad\qquad\\
&&\mbox{}+\,\bigl(y(b)-y(\tilde{b})+y(d)-y(\tilde{d})\bigr)\,
s(u\mi v)\,
s(u\pl v\mi\l)\,\tilde{g}(u)\,\tilde{g}(v)\bigr)\;.\end{eqnarray*}
The general solution of $\E_a(\tilde{b},\tilde{d})=0$ is 
\[\tilde{g}(u)=\tilde{x}_1\,f(u)\;,\qquad
\tilde{h}(u)=\tilde{x}_2\,f(u)\]
where $\tilde{x}_1$, $\tilde{x}_2$ and $f$ are arbitrary.
The remaining cases of $\E_a(b,d)=0$ now imply that
\[y(b)\;=\;\left\{\begin{array}{@{\,}l@{\,,\quad}l}
\rule[-1.5ex]{0ex}{1.5ex}
\tilde{y}&b\in\nu(a)\\
\tilde{y}'&b\in\nu'(a)\end{array}\right.\]
where $\tilde{y}$ and $\tilde{y}'$ are arbitrary, and
$\nu(a)$ and $\nu'(a)$ are non-intersecting sets whose union is
$\NN(a)$ and which contain $\tilde{b}$ and $\tilde{d}$ respectively.
The previous case in which all $y(b)$ are equal is included if we 
also allow $\nu(a)=\NN(a)$, $\nu'(a)=\emptyset$.
This now leads to the general solution~\eqref{TLDBW},
where $x_1(a)=(\tilde{y}-\tilde{y}')\tilde{x}_1$, 
$x_2(a)=\tilde{x}_2$ and $f(a,u)=f(u)$, and concludes our proof.

The Temperley-Lieb diagonal boundary weights satisfy
the boundary initial condition and the boundary inversion relation, 
with $\chi$ and $\hat{\rho}$ given by~\eqref{DBIC} and~\eqref{DBIR},
and boundary crossing symmetry with 
\beq\eql{TLBCS}\eta_a(u)=\frac{s(2\l\mi2u)}{s(\l)}\;
\frac{f(a,\l\mi u)}{f(a,u)}\;.\eeq
If, in \eqref{TLDBW}, we set
$x_1(a)=0$, $x_2(a)=1$ and $f(a,u)=1$
for each $a$, then we obtain boundary weights whose boundary face transfer
matrix is the identity matrix. That the identity satisfies~\eqref{MBYBE} 
is equivalent to the commutation of the bulk face transfer
matrices~\eqref{FTMcomm}.

The diagonal boundary weights for the critical $A_L$ and 
$D_L$ models, obtained 
by taking $q\rightarrow0$ in~\eqref{ALDBW} and~\eqref{DLDBW}, 
match those for the Temperley-Lieb models with adjacency
graphs $A_L$ and $D_L$ for appropriate choices of the arbitrary parameters.
In particular, for $i=2$ in~\eqref{DLDBW} we should set
$x_1(L\mi3)=
\ttt(0)\,(\tttt(0)^2\tilde{x}_1+\ttt(0)^2\tilde{x}_2)/q^{\ssm{1\!/2}}$
and\rule[-1.5ex]{0ex}{1.5ex}
$x_2(L\mi3)=
\pm\tttt(0)\,(\ttt(0)^2\tilde{x}_1+\tttt(0)^2\tilde{x}_2)/q^{\ssm{1\!/2}}$,
with $\tilde{x}_1$, $\tilde{x}_2$ and the $+$ or $-$ 
arbitrary,\rule[-1.5ex]{0ex}{1.5ex}
which gives
\[\begin{array}{@{}c@{}}
\rule[-2ex]{0ex}{2ex}k_2(L\mi3,u)\;\rightarrow\;
-\,(\tilde{x}_1\pl\tilde{x}_2)^2\,\sin(u\mi\l)^2\,+\,
2\,(\tilde{x}_1\pl\tilde{x}_2)\,\tilde{x}_2\\
k_2(L\mi1,u)\;=\;k_2(L,-u)\;\rightarrow\;
-\,(\tilde{x}_1\pl\tilde{x}_2)^2\,\sin(u\mipl\l)^2\,+\,
2\,(\tilde{x}_1\pl\tilde{x}_2)\,\tilde{x}_2
\;.\end{array}\]
The Temperley-Lieb models with adjacency graphs $E_6$, $E_7$ and $E_8$
can be related to those with adjacency graphs $A_{11}$, $A_{17}$ and
$A_{29}$ respectively by intertwiner cells~\cite{DifZub90,PeaZho93}.  
These cells satisfy the intertwiner inversion relations
and the bulk weight intertwining relation.
However, due to the absence of certain symmetries in the $E$ graphs,
we find that the only diagonal boundary weights  which satisfy the boundary 
weight intertwining relation are those which effectively correspond
to the identity solution.

\sect{Dilute $A_L$ Models}
\sub{Bulk Weights}
We now consider the dilute $A_L$ models~\cite{WarNieSea92}.
There is one such model for each integer $L\geq2$.
The spins in this model take values from the set $\{1,2,\ldots,L\}$ and 
the adjacency graph is
\setlength{\unitlength}{9mm}\beq\eql{dAL}
\raisebox{-0.6\unitlength}[0.6\unitlength][0.6\unitlength]{
\text{2.1}{1.2}{0}{0.6}{l}{A'_L\;\;=}\begin{picture}(6,1.2)
\multiput(0,0.5)(3.5,0){2}{\line(1,0){2.5}}
\multiput(2.7,0.5)(0.4,0){2}{\line(1,0){0.2}}
\multiput(0,0.5)(1,0){2}{\spos{}{\bullet}}
\multiput(5,0.5)(1,0){2}{\spos{}{\bullet}}
\multiput(0,0.7)(1,0){2}{\circle{0.4}}
\multiput(5,0.7)(1,0){2}{\circle{0.4}}
\put(0,0.35){\spos{t}{1}}\put(1,0.35){\spos{t}{2}}
\put(5,0.35){\spos{t}{L\mi1}}\put(6,0.35){\spos{t}{L}}
\end{picture}\text{0.4}{1}{0.4}{0.3}{b}{.}}\eeq
The dilute $A_L$ bulk weights are 
\beq\eql{dALFW}\begin{array}{@{\protect\rule[-4.5ex]{0ex}{4.5ex}}l}
\displaystyle\W{a}{a}{a}{a}{u}=
\frac{\t(6\ll\mi u)\:\t(3\ll\pl u)}{\t(6\ll)\:\t(3\ll)}\quad-\\
\displaystyle\qquad\Biggl(\frac{S_{a+1}}{S_a}\:\frac{\tttt((2a\mi5)\ll)}
{\tttt((2a\pl1)\ll)}+\frac{S_{a-1}}{S_a}\:\frac{\tttt((2a\pl5)\ll)}
{\tttt((2a\mi1)\ll)}\Biggr)\frac{\t(u)\:\t(3\ll\mi u)}{\t(6\ll)\:\t(3\ll)}
\;,\qquad a=1,\ldots,L\\
\displaystyle\W{a}{a}{a}{a\plmi1}{u}=\W{a}{a\plmi1}{a}{a}{u}=
\frac{\t(3\ll\mi u)\:\tttt((\pm2a\pl1)\ll\mi u)}
{\t(3\ll)\:\tttt((\pm2a\pl1)\ll)}\;,\\
\displaystyle\W{a\plmi1}{a}{a}{a}{u}=\W{a}{a}{a\plmi1}{a}{u}=
\left(\frac{S_{a\pm1}}{S_a}\right)^{\!\!\ssm{1\!/2}}\:\frac{
\t(u)\:\tttt((\pm2a\mi2)\ll\pl u)}{\t(3\ll)\:\tttt((\pm2a\pl1)\ll)}\;,\\
\W{a}{a\plmi1}{a\plmi1}{a}{u}=\W{a}{a}{a\plmi1}{a\plmi1}{u}\\
\displaystyle\qquad\qquad\qquad\qquad\qquad=
\Biggl(\frac{\tttt((\pm2a\pl3)\ll)\:\tttt((\pm2a\mi1)\ll)}
{\tttt((\pm2a\pl1)\ll)^2}\Biggr)^{\!\!\ssm{1\!/2}}\:
\frac{\t(u)\:\t(3\ll\mi u)}{\t(2\ll)\:\t(3\ll)}\;,\\
\displaystyle\W{a\plmi1}{a}{a\plmi1}{a}{u}=
\frac{\t(3\ll\mi u)\:\t((\pm4a\pl2)\ll\pl u)}
{\t(3\ll)\:\t((\pm4a\pl2)\ll)}\quad+\\
\displaystyle\qquad\qquad\qquad\qquad\qquad\quad\frac{S_{a\pm1}}{S_a}\:
\frac{\t(u)\:\t((\pm4a\mi1)\ll\pl u)}
{\t(3\ll)\:\t((\pm4a\pl2)\ll)}
\;,\qquad\biggl\{\begin{array}{@{}l}+,\;a=1,\ldots, L\mi1\\
-,\;a=2,\ldots,L\end{array}\\
\displaystyle\W{a}{a\mipl1}{a}{a\plmi1}{u}=
\frac{\t(2\ll\mi u)\:\t(3\ll\mi u)}{\t(2\ll)\:\t(3\ll)}\;,\\
\displaystyle\W{a\mipl1}{a}{a\plmi1}{a}{u}=
-\Biggl(\frac{S_{a-1}\:S_{a+1}}{S_a^2}\Biggr)^{\!\!\ssm{1\!/2}}\:
\frac{\t(u)\:\t(\ll\mi u)}{\t(2\ll)\:\t(3\ll)}
\;,\qquad a=1,\ldots, L\mi2
\end{array}\eeq
where
\beq\eql{lamdA}\ll\;=\;\frac{L}{L+1}\:\frac{\pi}{4}\quad\mbox{or}\quad
\frac{L+2}{L+1}\:\frac{\pi}{4}\eeq
and
\beq S_a\;=\;(-1)^a\:\frac{\t(4a\ll)}{\tttt(2a\ll)}\;.\eeq
These weights satisfy the Yang-Baxter equation, the initial condition, 
reflection symmetry, crossing symmetry with crossing parameter $\l=3\ll$
and crossing factors $S_a$,
and the inversion relation with 
\beq\eql{drho}\rho(u)=\frac{\t(2\ll\mi u)\:\t(3\ll\mi u)\:
\t(2\ll\pl u)\:\t(3\ll\pl u)}{\t(2\ll)^2\:\t(3\ll)^2}\;.\eeq
Invariance under the symmetry transformation~\eqref{ALZS}
is satisfied for $L$ even, but not for $L$ odd (assuming $q\ne0$).

\sub{Diagonal Boundary Weights}
We have found that the general diagonal solution of the boundary
Yang-Baxter equation for the dilute $A_L$ models is
\beq\eql{dALDBW}\begin{array}{@{}l@{}}
\B{a}{a}{a}{u}=\makebox[0mm][l]{$
\th_{i(a)}(\frac{5\ll}{2}\mi u)\:\th_{i(a)}(\frac{3\ll}{2}\pl u)\:
\th_{j(a)}((2a\mi\frac{1}{2})\ll\pl u)\:
\th_{j(a)}((2a\pl\frac{1}{2})\ll\mi u)\:
f(a,u)\;,$}\\
\rule[-1.5ex]{0ex}{1.5ex}\hspace{137mm}\makebox[0mm][l]{$a=1,\ldots,L$}\\
\rule[-4.5ex]{0ex}{4.5ex}\B{a}{a\plmi1}{a}{u}=\makebox[117mm][l]{$
\th_{i(a)}(\frac{5\ll}{2}\mi u)\,\th_{i(a)}(\frac{3\ll}{2}\mi u)\,
\th_{j(a)}((2a\mi\frac{1}{2})\ll\mipl u)\,
\th_{j(a)}((2a\pl\frac{1}{2})\ll\mipl u)\,
f(a,u)\;,$}\\
\rule[-4.5ex]{0ex}{4.5ex}\B{a\plmi1}{a}{a}{u}=\B{a}{a}{a\plmi1}{u}=0
\;,\qquad\biggl\{\begin{array}{@{}l}+,\;a=1,\ldots, L\mi1\\
-,\;a=2,\ldots,L\end{array}\\
\B{a\mipl1}{a}{a\plmi1}{u}=0\;,\qquad a=2,\ldots,L\mi1
\end{array}\eeq
where, for each $a$, $(i(a),j(a))$ may be chosen arbitrarily
as $(1,4)$, $(2,3)$, $(3,2)$ or $(4,1)$ and $f$ is arbitrary.
These weights match those obtained in~\cite{BatFriZho96},
and their derivation is similar to that of the dilute 
Temperley-Lieb diagonal  boundary weights
which is given in the next section.
They satisfy the boundary initial condition and the
boundary inversion relation, with $\chi$ and $\hat{\rho}$ given
by~\eqref{DBIC} and~\eqref{DBIR}, and boundary crossing symmetry with
\beq\eql{dALBCS}\eta_a(u)=
\frac{\t(6\ll\mi2u)\:\t(2u\mi\ll)}{\t(2\ll)\:\t(3\ll)}\;
\frac{f(a,3\ll\mi u)}{f(a,u)}\;.\eeq
For $L$ even, the weights can be made invariant under the 
transformation~\eqref{ALZS} for appropriate 
choices of the arbitrary parameters.
However, this is not possible for $L$ odd (assuming $q\ne0$, $f(a,u)\ne0$). 

\sect{Dilute Temperley-Lieb Models}
\sub{Bulk Weights}
We now consider the dilute Temperley-Lieb 
models~\cite{Roc92,WarNieSea92,WarNie93}.
The adjacency graph $\G$ for these models can be any finite, connected
graph which has only bidirectional, single bonds and in which each node is
connected to itself.

The dilute Temperley-Lieb bulk weights are
\beqa\eql{DTLFW}
\lefteqn{\rule[-3.5ex]{0ex}{3.5ex}\W{a}{b}{c}{d}{u}\;=\;
\rho_1(u)\,\delta_{abcd}\;+\;
\rho_2(u)\,\delta_{abc}\,\tilde{A}_{ad}\;+\;
\rho_3(u)\,\delta_{acd}\,\tilde{A}_{ab}\;+}\\
&&\!\rule[-3ex]{0ex}{3ex}\left(\frac{S_a}{S_b}\right)^{\!\!\ssm{1\!/2}}\!
\rho_4(u)\,\delta_{bcd}\,\tilde{A}_{ab}\;+\;
\left(\frac{S_c}{S_a}\right)^{\!\!\ssm{1\!/2}}\!
\rho_5(u)\,\delta_{abd}\,\tilde{A}_{ac}\;+\;
\rho_6(u)\,\delta_{ab}\,\delta_{cd}\,\tilde{A}_{ac}\;+\;
\rho_7(u)\,\delta_{ad}\,\delta_{bc}\,\tilde{A}_{ab}\;+\noeqno
&&\hspace{75mm}
\rho_8(u)\,\delta_{ac}\,\tilde{A}_{ab}\,\tilde{A}_{ad}\;+\;
\left(\frac{S_a\,S_c}{S_b\,S_d}\right)^{\!\!\ssm{1\!/2}}\!
\rho_9(u)\,\delta_{bd}\,\tilde{A}_{ab}\,\tilde{A}_{bc}\nonumber\eeqa
where $\delta_{a_1\ldots a_m}=\prod_{j=1}^{m-1}\delta_{a_ja_{j+1}}$ and
\beq\begin{array}{@{}l@{\quad\qquad}l@{}}\rule[-3.5ex]{0ex}{3.5ex}
\displaystyle\rho_1(u)\;=\;1+\frac{\sin\!u\,\sin(3\ll\mi u)}{\sin\!2\ll\,
\sin\!3\ll}&
\displaystyle\rho_2(u)\;=\;\rho_3(u)\;=\;
\frac{\sin(3\ll\mi u)}{\sin\!3\ll}\\
\rule[-3.5ex]{0ex}{3.5ex}\displaystyle
\rho_4(u)\;=\;\rho_5(u)\;=\;\frac{\sin\!u}{\sin\!3\ll}&\displaystyle
\rho_6(u)\;=\;\rho_7(u)\;=\;
\frac{\sin\!u\,\sin(3\ll\mi u)}{\sin\!2\ll\,\sin\!3\ll}\\
\rule[-3.5ex]{0ex}{3.5ex}\displaystyle
\rho_8(u)\;=\;
\frac{\sin(2\ll\mi u)\,\sin(3\ll\mi u)}{\sin\!2\ll\,\sin\!3\ll}&
\displaystyle\rho_9(u)\;=\;
-\frac{\sin\!u\,\sin(\ll\mi u)}{\sin\!2\ll\,\sin\!3\ll}\;.
\end{array}\eeq
Furthermore, $\tilde{A}=A-I$ is the adjacency 
matrix of the graph $\tilde{\G}$
obtained from $\G$ by removing the bonds connecting each node to itself,
$\Lambda$ is its maximum eigenvalue, $(S_1,S_2,\ldots)$ is the
associated eigenvector with all positive entries, 
and $\ll$ is any solution of
\beq\eql{lamDTL}\Lambda\;=\;-2\cos\!4\ll\;.\eeq
These weights satisfy the Yang-Baxter equation, the initial condition, 
reflection symmetry, crossing symmetry with crossing 
parameter~$\l=3\ll$ and crossing factors $S_a$,
and the inversion relation with 
\beq\eql{DTLrho}\rho(u)\;=\;\frac{\sin(2\ll\mi u)\:\sin(3\ll\mi u)\:
\sin(2\ll\pl u)\:\sin(3\ll\pl u)}{\sin^2\!2\ll\:\sin^2\!3\ll}\;.\eeq
{}From~\eqref{DTLFW} and~\eqref{TLS}, we see that
the bulk weights for the dilute Temperley-Lieb models
with adjacency graphs $A'_L$ match those for the 
critical dilute $A_L$ models, obtained by
taking $q\rightarrow0$ in~\eqref{dALFW}.

The dilute Temperley-Lieb bulk face transfer matrices can be expressed in 
terms of matrices $e^1_j$,\ldots,$e^9_j$ defined by~\eqref{DTLFW} and
\beq\eql{DTLFTM}X_j(u)\;=\;\sum_{n=1}^{9}\rho_n(u)\;e^n_j\;.\eeq
We then find that the matrices $e^n_j$ form a representation of the 
dilute Temperley-Lieb algebra~\cite{GriPea93} and that the Yang-Baxter 
equation~\eqref{MYBE} is satisfied through the 
relations of this algebra alone.
We also note that, in contrast with the Temperley-Lieb bulk face transfer
matrices~\eqref{TLFTM}, the dilute Temperley-Lieb bulk face transfer
matrices~\eqref{DTLFTM} do not commute.

\sub{Diagonal Boundary Weights}
Diagonal boundary weights for the dilute Temperley-Lieb models are
\beq\eql{DTLDBW}\B{a}{b}{c}{u}\;=\;\left\{
\begin{array}{@{}l@{\,,\quad}l}\rule[-2ex]{0ex}{2ex}
\sin(\xi(a)\mi\frac{\ll}{2}\pl u)\:\sin(\xi(a)\pl\frac{\ll}{2}\mi u)
\;f(a,u)\;\delta_{ac}&b=a\\\rule[-2ex]{0ex}{2ex}
\sin(\xi(a)\mi\frac{\ll}{2}\pl u)\:\sin(\xi(a)\pl\frac{\ll}{2}\pl u)
\;f(a,u)\;\delta_{ac}&b\in\nu(a)\\\rule{0ex}{0ex}
\sin(\xi(a)\mi\frac{\ll}{2}\mi u)\:\sin(\xi(a)\pl\frac{\ll}{2}\mi u)
\;f(a,u)\;\delta_{ac}&b\in\nu'(a)\end{array}\right.\eeq
where $f$ is arbitrary and, for each $a$,
$\nu(a)$ and $\nu'(a)$ are any non-intersecting
sets whose union is the set of neighbours of $a$ on $\tilde{\G}$
and $\xi(a)$ is any solution of
\beq\tan2\xi(a)\;=\;\frac{\sin\!4\ll}{\displaystyle\cos\!4\ll\,+
\!\!\sum_{d\,\in\,\nu(a)}\!\!\!S_d/S_a}\eeq
We now prove that these boundary weights represent the general
diagonal solution of the boundary Yang-Baxter equation for the 
dilute Temperley-Lieb models.
Having set $\smB{a}{b}{c}{u}=\BBB_a(b,u)\,\delta_{ac}$,
we find, using the dilute Temperley Lieb bulk weights, that
the only classes of spin assignments in~\eqref{BYBE}
which lead to non-trivial equations are\rule[-1.5ex]{0ex}{1.5ex}
$a=d=e$ and $b=c$ with $b\in\NN(a)$,
$a=c=d=e$ with $b\in\NN(a)$, and $a=c=e$\rule[-1.5ex]{0ex}{1.5ex} 
with $b\in\NN(a)$ and $d\in\NN(a)$, where
$\NN(a)$ is the set of neighbours of $a$ on $\tilde{\G}$.
These give, respectively,
\rnc{\theequation}{\arabic{section}.\arabic{equation}\alph{subeq}}
\setcounter{subeq}{1}\beq\eql{DTLBYBE1}
\begin{array}{@{}r@{}}\rule[-2ex]{0ex}{2ex}
0\;=\;\E^1_a(b)\;=\;\sin(u\mi v)\,(\BBB_a(a,u)\,\BBB_a(a,v)-
\BBB_a(b,u)\,\BBB_a(b,v))\qquad\\
+\;\sin(u\pl v)\,(\BBB_a(b,u)\,\BBB_a(a,v)-\BBB_a(a,u)\,\BBB_a(b,v))\;,
\end{array}\eeq
\beqa\rule[-2ex]{0ex}{2ex}\lefteqn{
0\;=\;\E^2_a(b)\;=\;\rho_4(u\mi v)\,\rho_1(u\pl v)\,
\BBB_a(a,u)\,\BBB_a(a,v)\;-\;
\rho_1(u\mi v)\,\rho_4(u\pl v)\,\BBB_a(a,u)\,\BBB_a(b,v)}\noeqno
\rule[-1.5ex]{0ex}{1.5ex}&&+\;
\rho_8(u\mi v)\,\rho_4(u\pl v)\,\BBB_a(b,u)\,\BBB_a(a,v)\;-\;
\rho_4(u\mi v)\,\rho_8(u\pl v)\,\BBB_a(b,u)\,\BBB_a(b,v)
\stepcounter{subeq}\addtocounter{equation}{-1}\eql{DTLBYBE2}\\
&&+\!\!\!\sum_{c\,\in\,\NN(a)}\!S_c\,\BBB_a(c,u)/S_a\,
(\rho_9(u\mi v)\,\rho_4(u\pl v)\,\BBB_a(a,v)-
\rho_4(u\mi v)\,\rho_9(u\pl v)\,\BBB_a(b,v))\;,\qquad\qquad\nonumber\eeqa
and
\beqa\rule[-2ex]{0ex}{2ex}\lefteqn{0\;=\;\E^3_a(b,d)\;=\;
\rho_9(u\mi v)\,\rho_8(u\pl v)\,
(\BBB_a(b,u)\,\BBB_a(b,v)-\BBB_a(d,u)\,\BBB_a(d,v))\;+}\noeqno
\rule[-2ex]{0ex}{2ex}&&\rho_8(u\mi v)\,\rho_9(u\pl v)\,
(\BBB_a(d,u)\,\BBB_a(b,v)-\BBB_a(b,u)\,\BBB_a(d,v))\;+
\stepcounter{subeq}\addtocounter{equation}{-1}\eql{DTLBYBE3}\\
\rule[-2ex]{0ex}{2ex}&&
\Bigl(\rho_4(u\mi v)\,\rho_4(u\pl v)\,\BBB_a(a,u)+
\rho_9(u\mi v)\,\rho_9(u\pl v)
\!\!\sum_{c\,\in\,\NN(a)}\!\!\!S_c\,\BBB_a(c,u)/S_a\Bigr)
(\BBB_a(b,v)-\BBB_a(d,v)).\nonumber\eeqa
We shall from now on treat $a$ as fixed.
If $a$ has $n$ neighbours on $\tilde{\G}$
then~\eqref{DTLBYBE1} and~\eqref{DTLBYBE2} each provide
$n$ equations and~\eqref{DTLBYBE3} provides $n(n\mi 1)/2$ 
equations for the $n+1$ boundary weights,
$\BBB_a(a,u)$ and $\BBB_a(b,u)$ with $b\in\NN(a)$. 
\rnc{\theequation}{\arabic{section}.\arabic{equation}}

Using a method similar to that for solving~\eqref{AIBYBE},
we find that the general solution of a single case 
of~\eqref{DTLBYBE1} can be written as
\[\begin{array}{c}\rule[-2ex]{0ex}{2ex}
\BBB_a(a,u)\,=\,(x_1\cos(u\mi\chi)+x_2\sin(u\mi\chi))\,f(u)\\
\rule{0ex}{0ex}\BBB_a(b,u)\,=\,
(x_1\cos(u\pl\chi)-x_2\sin(u\pl\chi))\,f(u)\end{array}\]
where $x_1$, $x_2$ and $f$\rule[-1ex]{0ex}{1ex} 
are arbitrary and $\chi$ may be set to any fixed value, which here 
we shall take as $\chi=\ll/2$.
Therefore, the general solution of the system of 
equations~\eqref{DTLBYBE1} is
\newpage
\beq\eql{DTLsol}\BBB_a(a,u)\,=\!\!
\prod_{c\,\in\,\NN(a)}\!\!\textstyle
(x_1(c)\cos(u\mi\frac{\ll}{2})+x_2(c)\sin(u\mi\frac{\ll}{2}))\,f(u)
\hspace{56mm}\eeq
\[\BBB_a(b,u)\,=\,{\textstyle
(x_1(b)\cos(u\pl\frac{\ll}{2})-x_2(b)\sin(u\pl\frac{\ll}{2}))}
\!\!\!\!\!\prod_{c\,\in\,\NN(a)-\{b\}}\!\!\!\!\!\!\textstyle
(x_1(c)\cos(u\mi\frac{\ll}{2})+x_2(c)\sin(u\mi\frac{\ll}{2}))\,
f(u)\]
where $x_1$, $x_2$ and $f$ are arbitrary.

We also observe that
\begin{eqnarray*}\rule[-1.5ex]{0ex}{1.5ex}\lefteqn{0\;=\;
\sin(u\mi v\mi\ll)\,(\E^2_a(b)\;-\;\E^2_a(d))\;+\;
\sin\!2\ll\;\E^3_a(b,d)\;=}\qquad\qquad\\
&&\sin(u\pl v)\,\sin(u\mi v\mi2\ll)\sin(u\mi v\mi3\ll)/
(\sin\!2\ll\,\sin^2\!3\ll)\;
\E^4_a(b,d)\end{eqnarray*}
where
\begin{eqnarray*}\rule[-1.5ex]{0ex}{1.5ex}
\E^4_a(b,d)&=&
\sin(u\mi v\pl\ll)\,\BBB_a(a,u)\,(\BBB_a(b,v)-\BBB_a(d,v))\\
\rule[-1.5ex]{0ex}{1.5ex}&&+\;
\sin(u\mi v\mi\ll)\,(\BBB_a(b,u)-\BBB_a(d,u))\,\BBB_a(a,v)\\
&&-\;\sin(u\pl v\mi\ll)\,(\BBB_a(b,u)\,\BBB_a(d,v)-
\BBB_a(d,u)\,\BBB_a(b,v))\;.\end{eqnarray*}
Using~\eqref{DTLsol}, we now find that
\[\begin{array}{@{}r@{}}\displaystyle
\rule[-4.5ex]{0ex}{4.5ex}0\;=\;\E^4_a(b,d)\;=
\!\!\!\!\!\!\!\prod_{c\,\in\,\NN(a)-\{b,d\}}\!\!\!\!\!\!\!
\textstyle(x_1(c)\cos(u\mi\frac{\ll}{2})+x_2(c)\sin(u\mi\frac{\ll}{2}))
(x_1(c)\cos(v\mi\frac{\ll}{2})+x_2(c)\sin(v\mi\frac{\ll}{2}))\\
\times\;\bigl(x_1(b)\,x_2(d)-x_1(d)\,x_2(b)\bigr)\,
\bigl(x_1(b)\,x_2(d)+x_1(d)\,x_2(b)\bigr)\,
\sin\!2u\,\sin\!2v\,\sin(u\mi v)\,f(u)\,f(v)\;.\end{array}\]
The general solution of this system of equations is
\[x_1(b)\;=\;\tilde{x}_1\:y(b)\;,\qquad
x_2(b)\;=\;\left\{\begin{array}{@{\,}r@{\,,\quad}l}
\rule[-1.5ex]{0ex}{1.5ex}
\tilde{x}_2\:y(b)&b\in\nu(a)\\
-\,\tilde{x}_2\:y(b)&b\in\nu'(a)\end{array}\right.\]
where $\tilde{x}_1$, $\tilde{x}_2$ and $y$ are arbitrary and
$\nu(a)$ and $\nu'(a)$ are any non-intersecting sets whose union is
$\NN(a)$.
This gives
\[\begin{array}{@{}l@{}}
\BBB_a(a,u)\;=\;\textstyle
(\tilde{x}_1\cos(u\mi\frac{\ll}{2})+
\tilde{x}_2\sin(u\mi\frac{\ll}{2}))\,
(\tilde{x}_1\cos(u\mi\frac{\ll}{2})-
\tilde{x}_2\sin(u\mi\frac{\ll}{2}))
\,\tilde{f}(u)\rule[-2.5ex]{0ex}{2.5ex}\\
\BBB_a(b,u)\;=\;\left\{\begin{array}{@{}l@{\,,\;\:}l}
\rule[-2ex]{0ex}{2ex}
(\tilde{x}_1\cos(u\pl\frac{\ll}{2})-
\tilde{x}_2\sin(u\pl\frac{\ll}{2}))\,
(\tilde{x}_1\cos(u\mi\frac{\ll}{2})-
\tilde{x}_2\sin(u\mi\frac{\ll}{2}))
\,\tilde{f}(u)&b\in\nu(a)\\\rule{0ex}{0ex}
(\tilde{x}_1\cos(u\pl\frac{\ll}{2})+
\tilde{x}_2\sin(u\pl\frac{\ll}{2}))\,
(\tilde{x}_1\cos(u\mi\frac{\ll}{2})+
\tilde{x}_2\sin(u\mi\frac{\ll}{2}))
\,\tilde{f}(u)&b\in\nu'(a)\end{array}\right.\end{array}\]
where
\[\textstyle\tilde{f}(u)\,=\,
(\tilde{x}_1\cos(u\mi\frac{\ll}{2})+
\tilde{x}_2\sin(u\mi\frac{\ll}{2}))^{|\nu(a)|-1}\,
(\tilde{x}_1\cos(u\mi\frac{\ll}{2})-
\tilde{x}_2\sin(u\mi\frac{\ll}{2}))^{|\nu'(a)|-1}
\!\!{\displaystyle\prod_{c\,\in\,\NN(a)}}\!\!\!y(c)\:f(u).\]
We now see that~\eqref{DTLBYBE3} is automatically satisfied for
$b,d\in\nu(a)$, or $b,d\in\nu'(a)$, while for 
$b\in\nu(a)$ and $d\in\nu'(a)$ we have
\begin{eqnarray*}\rule[-0.5ex]{0ex}{0.5ex}
\lefteqn{0\;=\;\E^3_a(b,d)\;=\;
2\sin\!2u\,\sin\!2v\,\sin(u\mi v)\,\sin(u\pl v)\,
\sin(u\mi v\mi\ll)\,\sin(u\pl v\mi\ll)/(\sin\!2\ll\,\sin\!3\ll)^2}\\
\rule[-2ex]{0ex}{2ex}&&\hspace{132mm}
\times\;\tilde{f}(u)\,\tilde{f}(v)\;\P\\
\rule[-1.5ex]{0ex}{1.5ex}\lefteqn{\textstyle0\;=\;\E^2_a(b)\;=\;
(\tilde{x}_1\cos(u\mi\frac{\ll}{2})-\tilde{x}_2\sin(u\mi\frac{\ll}{2}))\,
(\tilde{x}_1\cos(v\mi\frac{\ll}{2})-\tilde{x}_2\sin(v\mi\frac{\ll}{2}))}\\
\rule[-2.5ex]{0ex}{2.5ex}&&\hspace{46mm}
\times\;\sin\!2u\,\sin\!2v\,\sin(u\mi v)\,\sin(u\pl v)/
(\sin\!2\ll\,\sin^2\!3\ll)\,\tilde{f}(u)\,\tilde{f}(v)\;\P\\
\rule[-1.5ex]{0ex}{1.5ex}\lefteqn{\textstyle0\;=\;\E^2_a(d)\;=\;
(\tilde{x}_1\cos(u\mi\frac{\ll}{2})+\tilde{x}_2\sin(u\mi\frac{\ll}{2}))\,
(\tilde{x}_1\cos(v\mi\frac{\ll}{2})+\tilde{x}_2\sin(v\mi\frac{\ll}{2}))}\\
&&\hspace{46mm}\times\;\sin\!2u\,\sin\!2v\,\sin(u\mi v)\,\sin(u\pl v)/
(\sin\!2\ll\,\sin^2\!3\ll)\,\tilde{f}(u)\,\tilde{f}(v)\;\P
\end{eqnarray*}
where 
\[\P\;=\;\sin\!4\ll\,\bigl(\tilde{x}_1^2\,-\,
\tilde{x}_2^2\bigr)\:-\:
2\,\Bigl(\cos\!4\ll\,+\!\!\sum_{d\,\in\,\nu(a)}\!\!S_d/S_a\Bigr)\,
\tilde{x}_1\,\tilde{x}_2\;,\]
and we have used the eigenvector equation
\[\sum_{b\,\in\,\NN(a)}S_b/S_a\;=\;-2\,\cos\!4\ll\;.\]
We must therefore set $\P=0$, the general solution of which is
\[\tilde{x}_1\;=\;z\,\sin\xi\,,\qquad\tilde{x}_2\;=\;-\,z\,\cos\xi\]
where $z$ is arbitrary and $\xi$ is any solution of
\[\tan2\xi\;=\;\frac{\sin\!4\ll}{\displaystyle\cos\!4\ll\,+
\!\!\sum_{d\,\in\,\nu(a)}\!\!\!S_d/S_a}
\;=\;-\,\frac{\sin\!4\ll}{\displaystyle\cos\!4\ll\,+
\!\!\sum_{d\,\in\,\nu'(a)}\!\!\!S_d/S_a}\;.\]
This now leads to the general solution~\eqref{DTLDBW},
where $\xi(a)=\xi$ and $f(a,u)=z^2\,\tilde{f}(u)$, and concludes our proof.

The dilute Temperley-Lieb diagonal boundary weights
satisfy the boundary initial condition and the
boundary inversion relation, with $\chi$ and $\hat{\rho}$ given
by~\eqref{DBIC} and~\eqref{DBIR}, and boundary crossing symmetry with
\beq\eql{DTLBCS}\eta_a(u)=
\frac{\sin(6\ll\mi2u)\:\sin(2u\mi\ll)}{\sin\!2\ll\:\sin\!3\ll}\;
\frac{f(a,3\ll\mi u)}{f(a,u)}\;.\eeq
We also see that the diagonal boundary weights for the critical dilute $A_L$
models, obtained by replacing $f(a,u)$ by $f(a,u)/q^{\ssm{1\!/2}}$ and
taking \rule[-1.5ex]{0ex}{1.5ex}$q\rightarrow0$ in~\eqref{dALDBW},
match those for the dilute Temperley-Lieb models with adjacency 
graphs $A'_L$.

\rule[-1ex]{0ex}{1ex}
If, in \eqref{DTLDBW}, we set $\nu(a)$ equal to the set
of neighbours of $a$, $\nu'(a)=\emptyset$, $f(a,u)=1$
and $\xi(a)=-2\ll$ or $\xi(a)=-2\ll+\pi/2$, for all $a$, then
we obtain
\beq\eql{DTLABW}\B{a}{b}{c}{u}=\hat{\rho}_1(u)\,\delta_{abc}\;+\;
\hat{\rho}_2(u)\,\delta_{ac}\,\tilde{A}_{ab}\eeq
where 
\beq\begin{array}{@{}l@{\quad}l@{\qquad}l@{}}
\rule[-1.5ex]{0ex}{1.5ex}&\textstyle\hat{\rho}_1(u)\;=\;
\sin(\frac{5\ll}{2}\mi u)\,\sin(\frac{3\ll}{2}\pl u)\;,&
\hat{\rho}_2(u)\;=\;
\sin(\frac{5\ll}{2}\mi u)\,\sin(\frac{3\ll}{2}\mi u)\\
\mbox{or}&\textstyle\hat{\rho}_1(u)\;=\;
\cos(\frac{5\ll}{2}\mi u)\,\cos(\frac{3\ll}{2}\pl u)\;,&
\hat{\rho}_2(u)\;=\;
\cos(\frac{5\ll}{2}\mi u)\,\cos(\frac{3\ll}{2}\mi u)\;.\end{array}\eeq
We have found that the boundary face transfer matrices obtained 
from~\eqref{DTLABW} satisfy the boundary Yang-Baxter 
equation~\eqref{MBYBE} 
through the relations of the dilute Temperley-Lieb algebra alone.

\sect{Discussion}
We have obtained general solutions, mostly of the diagonal type, of
the boundary Yang-Baxter equation for a number of related
interaction-round-a-face models.  The boundary weights all involve
arbitrary parameters, some of which may take any complex value, while
others of which may take only finitely many values.

The solutions were derived by direct consideration of the relevant
equations for each model.  In some cases our solutions include
boundary weights which can also be obtained by indirect means.
These alternative methods include the consideration of algebraic
relations associated with the model, the construction of new weights
from known, simpler weights using fusion~\cite{BehPea96}, and the
generation of weights for one model from known weights for another
model using vertex-face intertwiners~\cite{FanHouShi95} or face-face
intertwiners.  Such means are useful for establishing the existence
of solutions and for efficiently deriving particular solutions which
are adequate for many purposes.  However it seems that a direct
approach is still needed in order to obtain general solutions, and
thereby to identify the exact number and nature of associated
arbitrary parameters.

\section*{Acknowledgement}
This work was supported by the Australian Research Council.

\end{document}